\documentclass[final,5p,times,twocolumn]{elsarticle}

\usepackage{graphicx}
\usepackage[update,prepend]{epstopdf}
\usepackage{amsmath}

\usepackage{amssymb}

\usepackage{lineno}

\usepackage{xspace}
\journal{Nucl. Instrum. Methods Phys. Res., Sect. A}
\newcommand{\dcube}{${\rm D}^3$-Micro\xspace}

\begin{document}
\begin{frontmatter}



\title{3-D Tracking of Nuclear Recoils in a Miniature Time Projection Chamber}

\author[]{S.~E.~Vahsen\corref{cor1}}
\cortext[cor1]{Corresponding author. Tel.: +1 808 956 2985.}
\ead{sevahsen@hawaii.edu}
\author{M.~T.~Hedges}
\author{I.~Jaegle} 
\author{S.~J.~Ross}
\author{I.~S.~Seong}
\author{T.~N.~Thorpe}
\author{J.~Yamaoka}
\address{University of Hawaii, 2505 Correa Road, Honolulu, HI 96822, USA}
\author{J.~A.~Kadyk}
\author{M.~Garcia-Sciveres}
\address{Lawrence Berkeley National Laboratory, 1 Cyclotron Road, Berkeley, CA 94720, USA}


\begin{abstract}
The three-dimensional (3-D) reconstruction of nuclear recoils is of interest for directional 
detection of fast neutrons and for direction-sensitive searches for weakly interacting massive particles 
(WIMPs), which may constitute the Dark Matter of the universe. We demonstrate this capability 
with a miniature gas target Time Projection Chamber (TPC) where the drift charge 
is avalanche-multiplied with Gas Electron Multipliers (GEMs) and detected with 
the ATLAS FE-I3 Pixel Application Specific Integrated Circuit (ASIC). We report on performance characterization of the detector, including measurements of gain, gain resolution, point resolution, diffusion, angular resolution, and energy resolution with low-energy x-rays, cosmic rays, and alpha particles, using 
the gases Ar:CO$_2$ (70:30) and He:CO$_2$ (70:30) at atmospheric pressure. We discuss the implications 
for future, larger directional neutron and Dark Matter detectors. With an eye to designing and 
selecting components for these, we generalize our results into analytical expressions for detector 
performance whenever possible. We conclude by demonstrating the 3-D directional detection of a fast neutron source.
\end{abstract}


\end{frontmatter}


\section{Introduction}
\label{introduction}
\begin{figure}
\centering
\includegraphics[width=8cm]{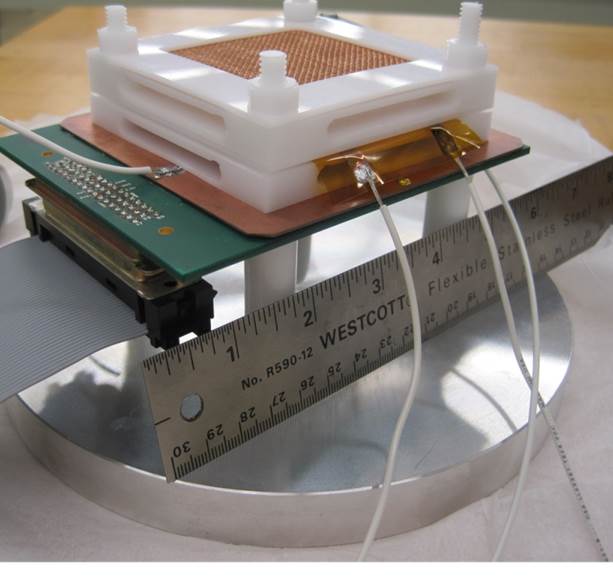}
\setlength{\abovecaptionskip}{0pt}
\caption{ \dcube prototype in the original configuration used for most studies presented here. The sensitive volume consists of a 9.2-mm vertical drift gap between the copper cathode (mesh visible on the top of the detector) and the top GEM (foil protruding on the right). 
}
\setlength{\belowcaptionskip}{0pt}
\label{detector}
\end{figure}
Time Projection Chambers \cite{Nygren:1978rx} with charge readout via micro-pattern 
gaseous detectors are the digital analog to bubble chambers, allowing 3-D reconstruction of ionization 
with many space points and great precision. Over the last decade, a number of studies have 
demonstrated impressive performance when reconstructing ionizing primary particles with such detectors \cite{Bellazzini:2004hr, Colas:2004ks, Kim:2008zzi}. Our group is investigating \cite{Seong_vci2013,Ross:2013bza,Yamaoka:2012ux,Vahsen:2014mca,Jaegle:2012sma} the detection of nuclear recoils resulting from the scattering 
of neutral primary particles, such as neutrons and, potentially, WIMPs. Neutron detectors 
with improved directional sensitivity are likely to find applications in particle physics, 
nuclear physics, homeland security, and neutron imaging. Directional searches for WIMP Dark Matter are 
sensitive to a unique signature, the 24-hour directional oscillation in the mean WIMP recoil direction 
due to the rotation of the earth \cite{Spergel:1987kx}. Observation of this signature would constitute a convincing detection of WIMPs by demonstrating the cosmological origin of the signal, and may be required to distinguish WIMP scattering from coherent neutrino scattering \cite{Grothaus:2014hja}.  A number
of  technological approaches are being explored \cite{Ahlen:2010ub,Daw:2010ud,Vahsen:2011qx,Santos:2011kf,Miuchi:2010hn,Naka:2011sf,Drukier:2012hj, Nygren:2013nda,Ahlen:2009ev}. An ideal directional WIMP detector, capable of excluding the isotropy of nuclear recoils in galactic coordinates with order ten signal events, would track nuclear recoils in 3-D, with low track energy threshold and head/tail recognition \cite{Green:2006cb}.
 The technology under study is a candidate for building such a detector. One obvious challenge
for gas-based WIMP searches is low target mass per unit volume. However, the proper metric for comparing
technologies is sensitivity per unit of cost. For high-resolution gas TPCs, the cost drivers are typically the 
readout plane and electronics. If the cost of these can be minimized, for instance by focusing the drift charge onto 
the detection plane \cite{Ross:2013bza}, large gas TPCs will be more competitive.

We report here on the performance 
of a miniature prototype, \dcube (Directional Dark Matter Detector - Micro), 
constructed at the University of Hawaii in 2010.  In that detector the TPC drift charge is multiplied 
with a double layer of Gas Electron Multipliers (GEMs) \cite{Sauli:1997qp} and 
detected with the ATLAS FE-I3 Pixel Application Specific Integrated Circuit (ASIC) \cite{Aad:2008zz}. 
The high double GEM gain (of order $10^4$), low pixel threshold (typically 2000-4000 e$^-$), and 
low pixel noise (typically 100-200 e$^-$) result in several attractive features, such as stable operation
with single electron efficiency near unity, self-triggered readout, and negligible rates of noise hits. In practical 
terms, this means that at high gain, essentially all primary ionization can be detected, so that the energy threshold 
is equal to the work function of the gas, typically about 30~eV. Therefore one can expect a large number of 
hits even for keV-scale tracks. It seems likely that these outstanding capabilities will enable reconstruction of tracks with the lowest track energy threshold possible in any detector of ionization. Due to the self-triggering capability of the pixel chip, the detector produces no data in the absence of ionization in the drift gap, greatly reducing the requirements on DAQ electronics. This is important in the context of scaling to larger detectors.
\begin{figure}
\centering
\includegraphics[width=8cm]{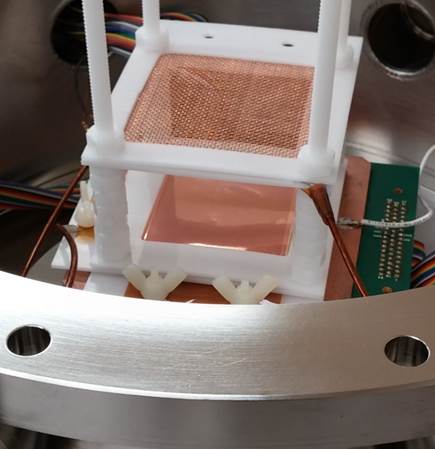}
\caption{ \dcube prototype in the test vessel, after the drift gap was increased to 45~mm, and the thickness of nearby Delrin (acetal) parts was reduced. These modifications were crucial for achieving a significant directional neutron signal.}
\label{detector2}
\end{figure}

All measurements presented here were carried out at atmospheric pressure (760 Torr). Initial measurements employed Ar:CO$_2$ (70:30), which is a commonly used detector gas, and allows a comparison with work by others. Later measurements were performed with He:CO$_2$ (70:30), which is more suitable for reconstructing fast neutron recoils. Helium is a good neutron target, since up to 64\% of the neutron energy can be transfered to a helium nucleus. Helium is also a good detection medium, since the low electron density results in small specific 
ionization, yielding longer recoil tracks. The CO$_2$ component improves detector performance and stability by reducing diffusion, improving quenching, and raising the electric field strength threshold for sparking.

\begin{figure}
\centering
\includegraphics[width=9cm, trim=1.2cm 21.3cm 13cm 1.0cm, clip=true,]{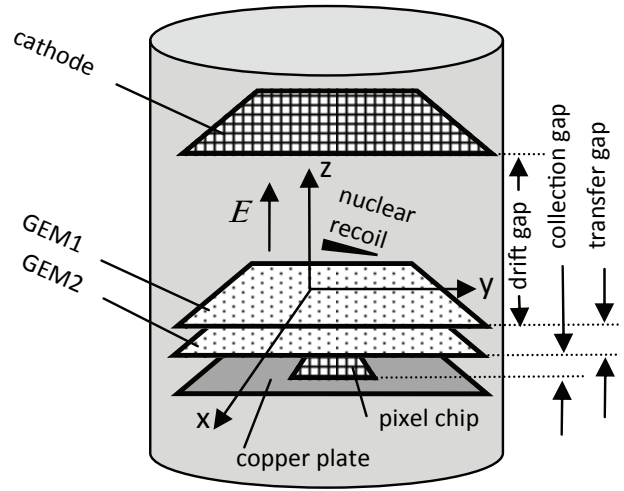}
\setlength{\abovecaptionskip}{0pt}
\caption{Schematic representation of the \dcube prototype. \label{fig:detectordrawing}}
\setlength{\belowcaptionskip}{0pt}
\end{figure}
\section{Detector and Principle of Operation}

The \dcube prototype consists of a Delrin (acetal) support structure, visible as white parts in Figs.~\ref{detector} and~\ref{detector2}, on which the different electrical components are mounted. The support structure resides inside a 25-liter stainless steel test vessel. Much of the detector design, shown schematically in Fig.~\ref{fig:detectordrawing}, is based on a previous prototype constructed at LBNL \cite{Kim:2008zzi}. The sensitive volume of the detector consists of a drift gap situated between a copper mesh and the upper surface of GEM1. For most measurements the drift gap was 9.2~mm. For the demonstration of neutron detection, higher detection efficiency was required, and the drift gap was increased to 45~mm, as shown in Fig.~\ref{detector2}. Ionizing radiation produces free electrons in this gap. These electrons then drift in a uniform electric field to a double GEM layer, where the electrons are avalanche multiplied, and finally the resulting avalanche charge is detected with an ATLAS FE-I3 pixel chip \cite{Aad:2008zz} operating in self-trigger mode (described below) and sampling at 40 MHz. The GEMs used are the standard CERN design with an active area of $5\times 5$~cm and 140 $\mu {\rm m}$ hole spacing \cite{Sauli:1997qp}. The pixel chip has an active area of $7.2 \times 8.0~{\rm mm}^2$, divided into 2880 pixels. Each individual $50 \times 400~\mu{\rm m}^2$ pixel contains an integrating amplifier, a discriminator, a shaper, and associated digital controls. Since most of the pixel chip surface is non-conductive, a pixelized metal layer was deposited \cite{Kim:2008zzi} onto the chip to increase the charge collection efficiency. (This did, however, not improve charge collection as intended -- see the section on energy resolution.) When charge is detected in the chip (at least one pixel detects charge above threshold), the self-trigger results in the output of a zero-suppressed digital serial stream that encodes the 2-D position, arrival time, and amount of charge collected, for each pixel above threshold within the next sixteen cycles of 25 ns each. The charge collected in each pixel is deduced from the time above threshold (ToT), which is measured with 7-bit precision. By using the known drift velocity in the drift gap, the timing information is converted into a (relative) third spatial coordinate, so that a 3-D image of ionization in the drift gap is obtained, as shown in Figs.~\ref{fig:cosmic}, \ref{fig:alpha}, and \ref{fig:recoil}. The pixel chip is glued to a circuit board and electrically connected with wirebonds, which are shielded against the electric field with a small metal overhang, as described in \cite{Kim:2008zzi}. In addition to the digital charge readout via the pixel chip, the area surrounding the chip is covered with a copper plate which is connected via a capacitor to an Endicott eV-5093 charge sensitive preamplifier. The amplifier output is fed through a Canberra AFT 2025 shaping amplifier into an Ortec EASY-MCA operating in pulseheight analyzer (PHA) mode. The PHA is used to measure gain and gain resolution of the double GEM.

For the studies described here, the test vessel was typically pumped down to $10^{-4}$~Torr, and then filled with the target gas under study. To ensure good gas purity, we usually performed multiple such pump-and-fill sequences before data taking, repeated the gain studies with and without gas flow during detector operation, and checked for long term gain stability, which was better than 2\% over several weeks. For studies with x-rays and alpha-particles, radioactive sources were placed inside the vessel. For neutron detection studies, we placed the source outside the vessel. We adjusted the double GEM gain and drift field depending on the energy scale and requirements of each particular study. The different settings are summarized in Table \ref{table:gain_and_efields}.

\begin{table*}
\begin{center}
    \begin{tabular}{|l|c|c|c|c|c|}
    \hline
    source of  & $V_{GEMS}$ & effective gain  & $E_{drift}$ & $E_{transfer}$ & $E_{collection}$\\ 
 	ionization & 	&	($10^3$) 	&	(kV/cm)	&	(kV/cm)	&	(kV/cm)	 \\\hline
    cosmic rays         	& 983  	& 41  	& 1.2  	   & 3.7  	& 2.7             \\ 
    alpha particles   	& 833   	& 3.2    	& 0.84    & 2.3   	& 3.1          \\ 
    neutron recoils   	& 806    	& 2.0		& 0.64    & 2.2   	& 3.0            \\ \hline
    \end{tabular}

\caption{Gain and electric field settings used for the different studies presented. $V_{GEMS}$ is the sum of the two, nearly equal, voltages across GEM1 and GEM2. \label{table:gain_and_efields}.}
\end{center}

\end{table*}

\section{Measurements of Gain and Gain Resolution\label{sec:gain}}
\begin{figure}
\centering
\includegraphics[width=8cm]{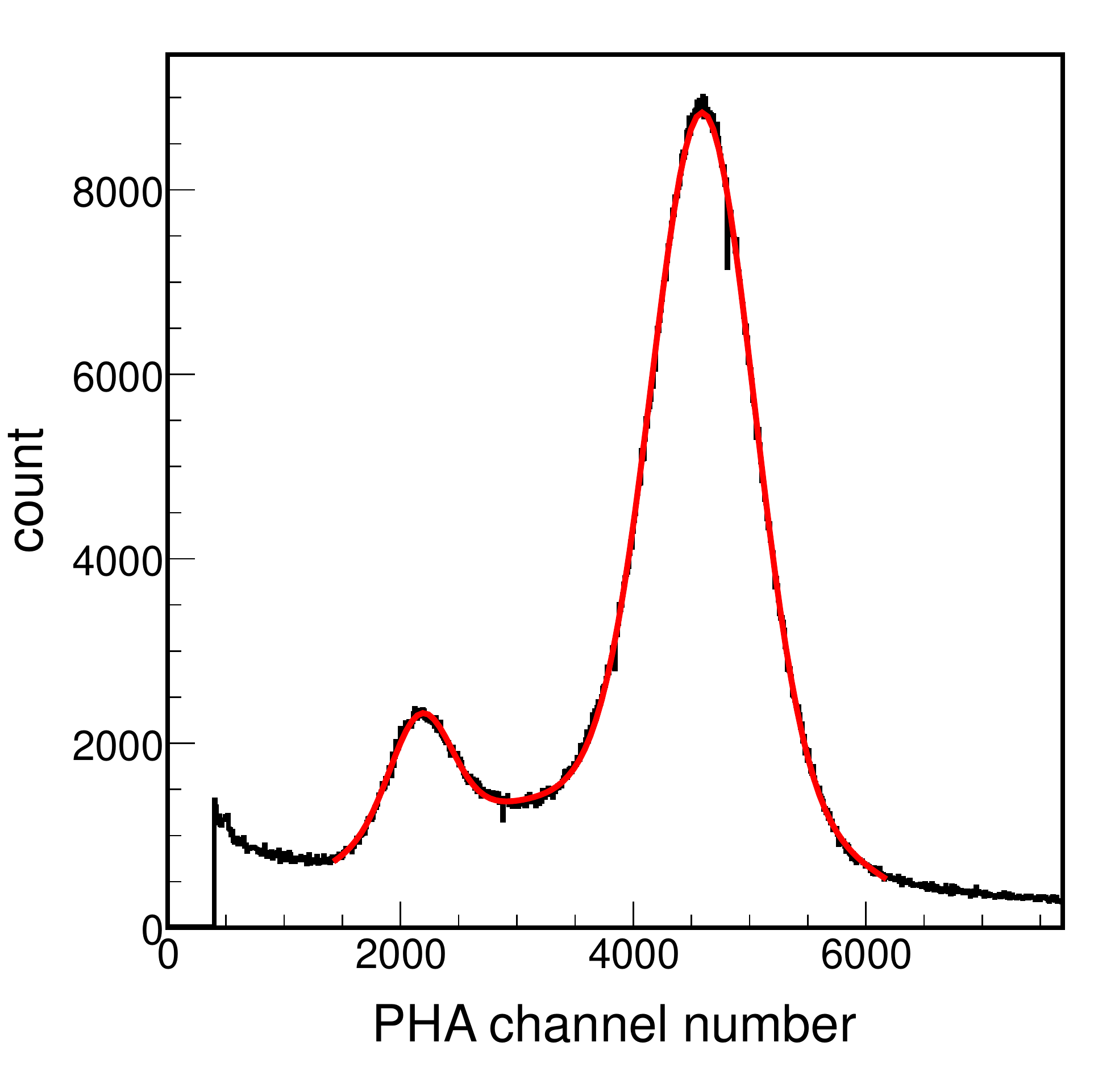}
\includegraphics[width=8cm]{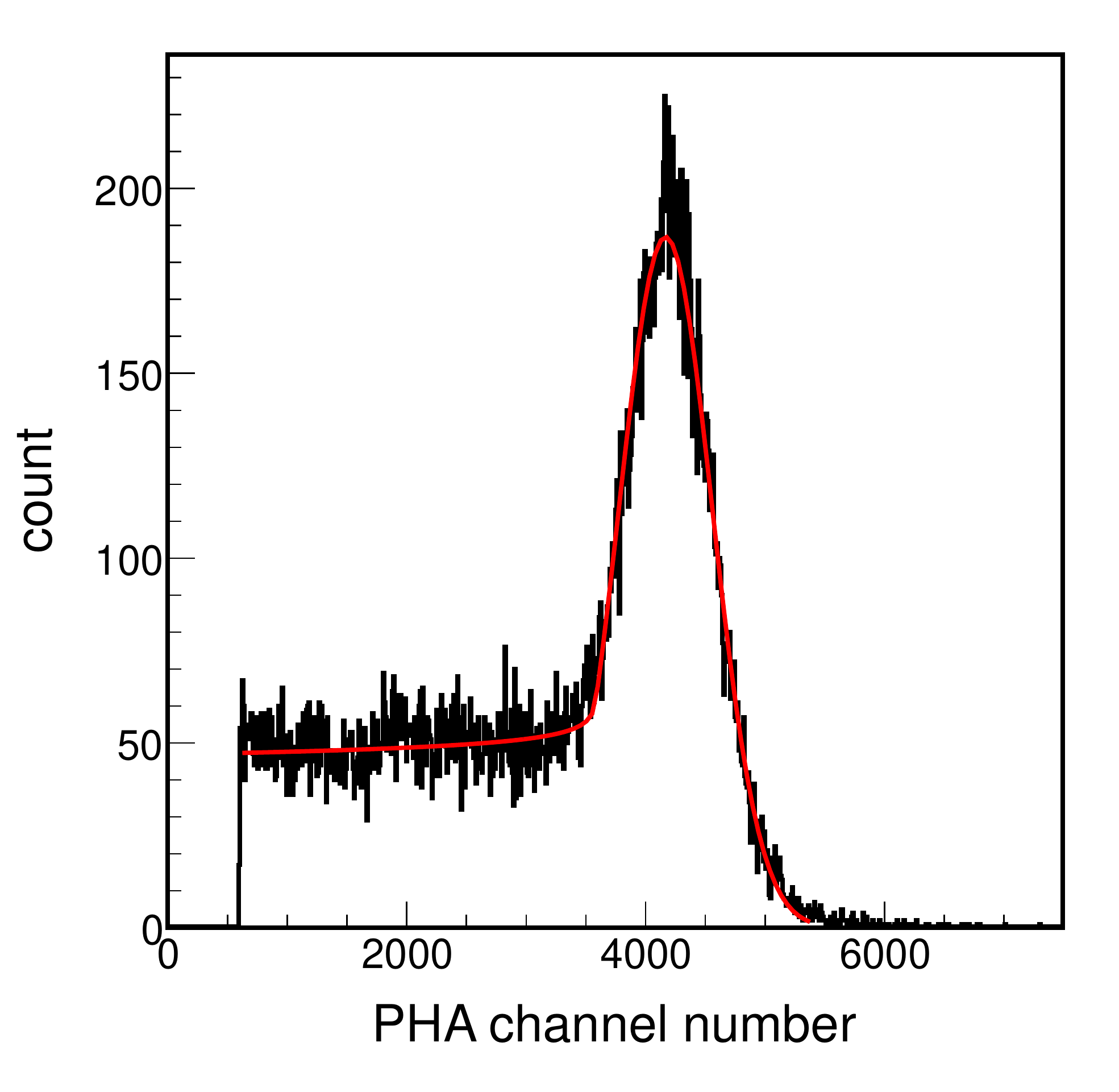}
\setlength{\abovecaptionskip}{0pt}
\caption{Pulseheight spectra recorded with an Fe-55 x-ray source using ArCO$_2$ (upper) and HeCO$_2$ (lower) gas and a double-GEM gain of $4\times10^4$. The black points show experimental data.  The smooth curves (red in online version) are the result of fits to the data, as described in the text. \label{fig:spectra}}
\setlength{\belowcaptionskip}{0pt}
\end{figure}

\begin{figure}
\centering
\includegraphics[width=8cm]{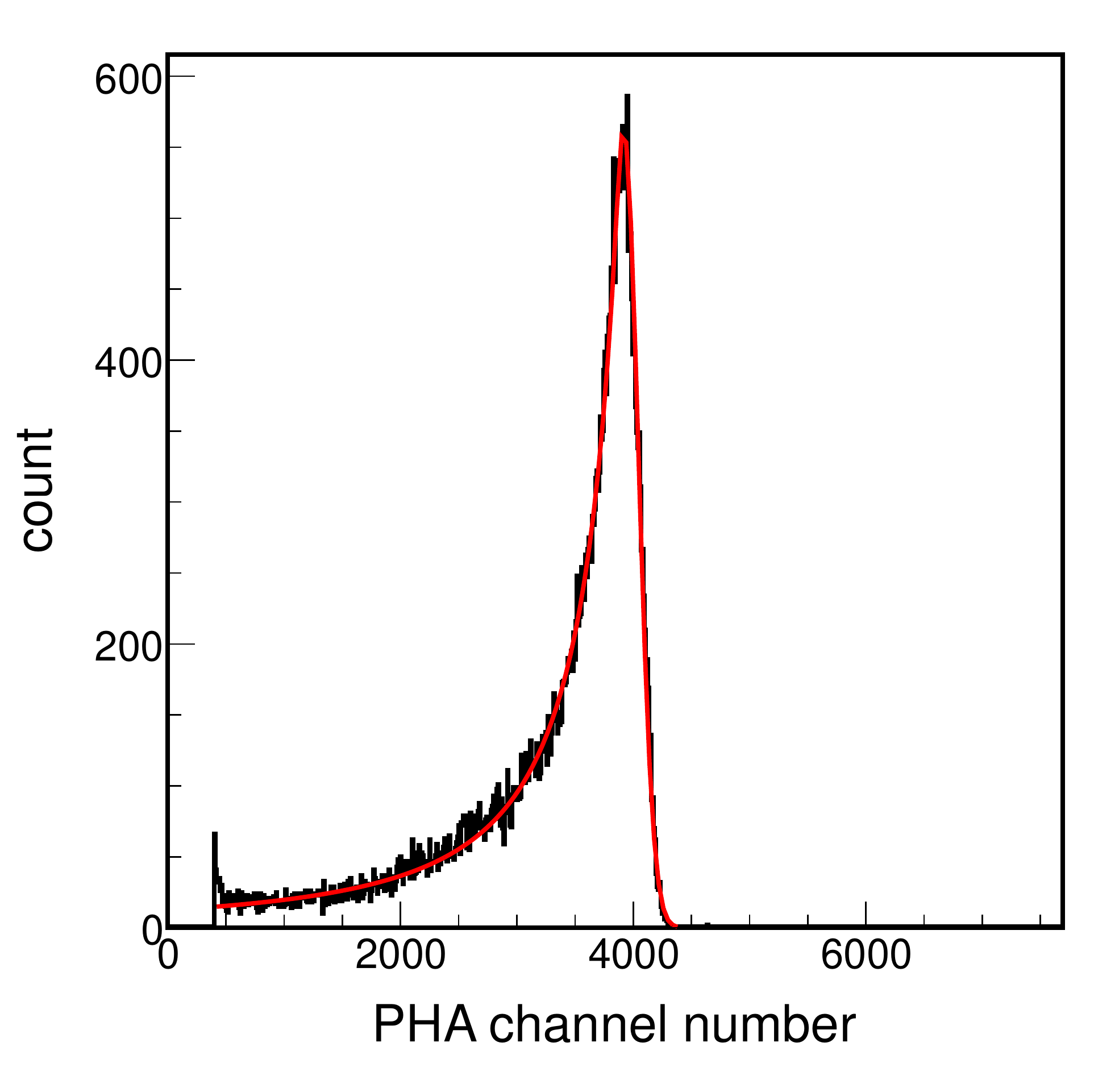}
\setlength{\abovecaptionskip}{0pt}
\caption{Pulseheight spectrum recorded with a Po-210 alpha source in ArCO$_2$ gas at a double-GEM gain of approximately 45.  The black points show experimental data. The smooth curve (red online) is the result of a fit to the data, as described in the text.
 \label{fig:alphaspectrum}}
\setlength{\belowcaptionskip}{0pt}
\end{figure}

\subsection{X-rays}

We measure the effective double-GEM gain and its resolution by placing an uncollimated Fe-55 5.9-keV x-ray source on top of the cathode mesh and observing the resulting pulse height spectra in the PHA. Two of the many spectra measured are shown in Fig.~\ref{fig:spectra}. We perform a $\chi^2$-minimization to the spectra, where the ArCO$_2$ x-ray conversion signal is modeled by the sum of two Gaussians (the argon main peak and escape peak). We perform each fit twice, once with the noise background in the PHA spectrum modeled by a polynomial of degree two, and once with the noise modeled by a power-law function. We take the mean fit parameters obtained with the two background models as our result, and the differences between results with the two models, in quadrature with fitting uncertainties, as systematic uncertainties. For HeCO$_2$ spectra we model the signal with a Crystal Ball function \cite{Oreglia:1980cs}, i.e. a Gaussian peak with a power-law low-end tail below an adjustable threshold, and obtain a good fit to the relevant part of the spectrum without including any additional background component. For both gases, the mean of the Gaussian is used to calculate a gain value, and the sigma of the Gaussian is used to calculate the gain resolution at that gain, as follows: We convert PHA channel numbers into detected charge values using the measured response of the PHA and amplifier chain. The latter was measured to be $0.91\pm0.13$~V/pC, using an injection capacitor and voltage pulse generator. The quoted uncertainty is systematic, limited by our measurement of the small injection capacitance. As a result, all effective gain measurements presented have a common, $14\%$ systematic uncertainty, which we do not include in error bars in any figures. We assume that the x-ray conversion yields 210 electrons for the main peak in ArCO$_2$, and 172 electrons for the peak in HeCO$_2$ \cite{Sharma:1998xw}. The ratio of detected charge to the x-ray conversion yield is taken to be the effective gain.
 
\subsection{Alpha Particles}
Since the gain resolution improves with increasing number of primary electrons, we also performed ArCO$_2$  gain and resolution measurements with a Po-210 5.3-MeV alpha particle source. The source was aimed horizontally so that the alpha particles would enter and stop in the drift gap above the copper pad that feeds the PHA. In this configuration we expect $4 \pm 1.3$~ MeV of primary ionization to contribute to the PHA signal. The large uncertainty on this initial energy stems from an uncertainty in the exact position of the source. As can be seen in Fig.~\ref{fig:alphaspectrum}, the resulting PHA spectra are much narrower than those at lower energies. We fit the PHA spectra with a Crystal Ball function, and extract the gain and its resolution as described above for x-rays. Note that this implies that the narrow upper tail of the pulseheight peak is used to measure the gain resolution. The broad, lower tail of the peak is dominated by alpha particles that are emitted at a small angle from the source, thus taking a slightly longer trajectory (and having lower energy) before reaching the drift gap, and depositing less ionization in the drift gap. Conversely, the upper tail of the peak pulseheight spectrum is due to alpha particles taking the shortest possible trajectory to reach the drift gap. There is thus a sharp upper limit for the deposited energy, and the width of the upper tail reflects the gain resolution.

\begin{figure}
\centering
\includegraphics[width=8cm]{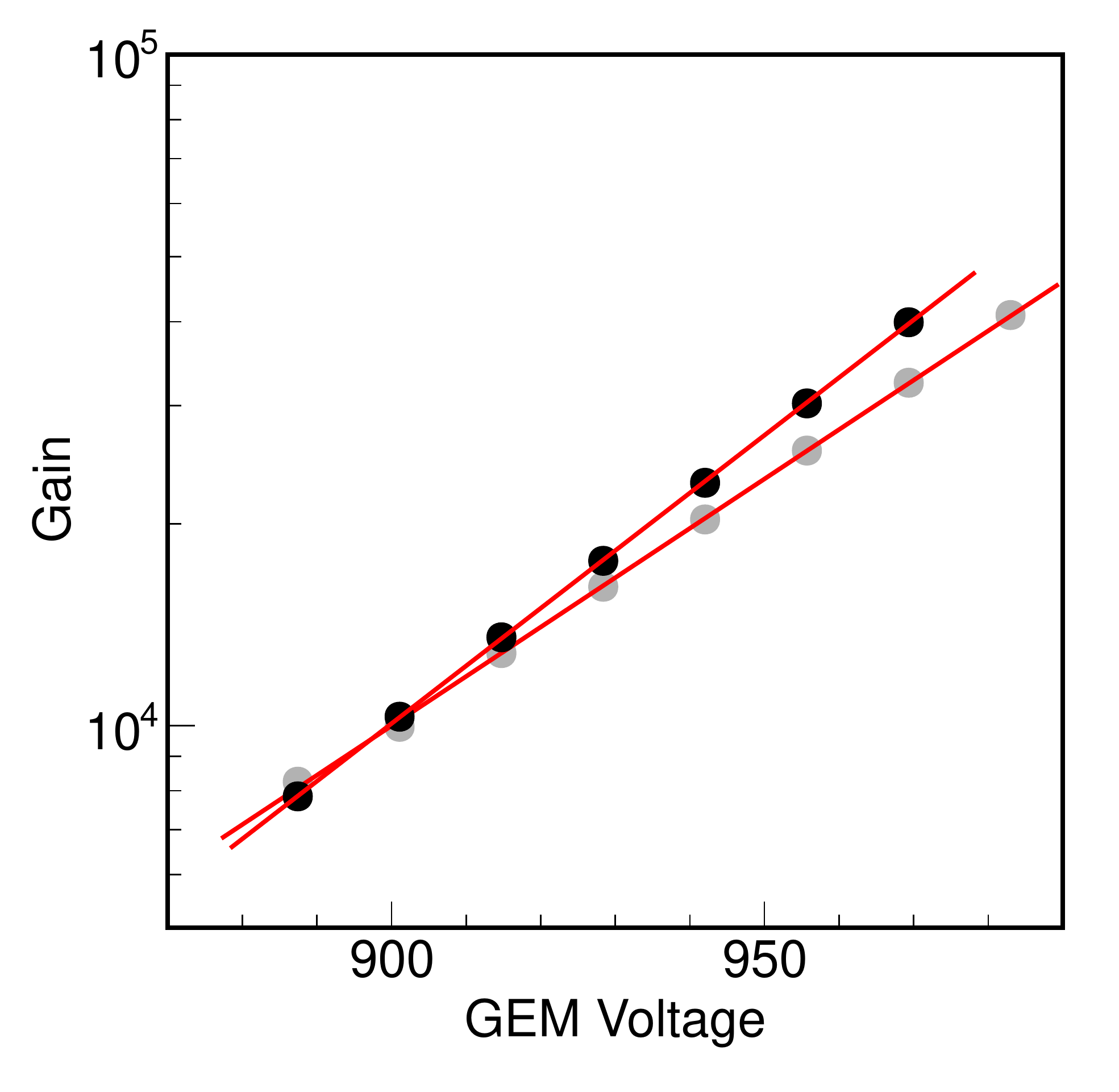}\\\includegraphics[width=8cm]{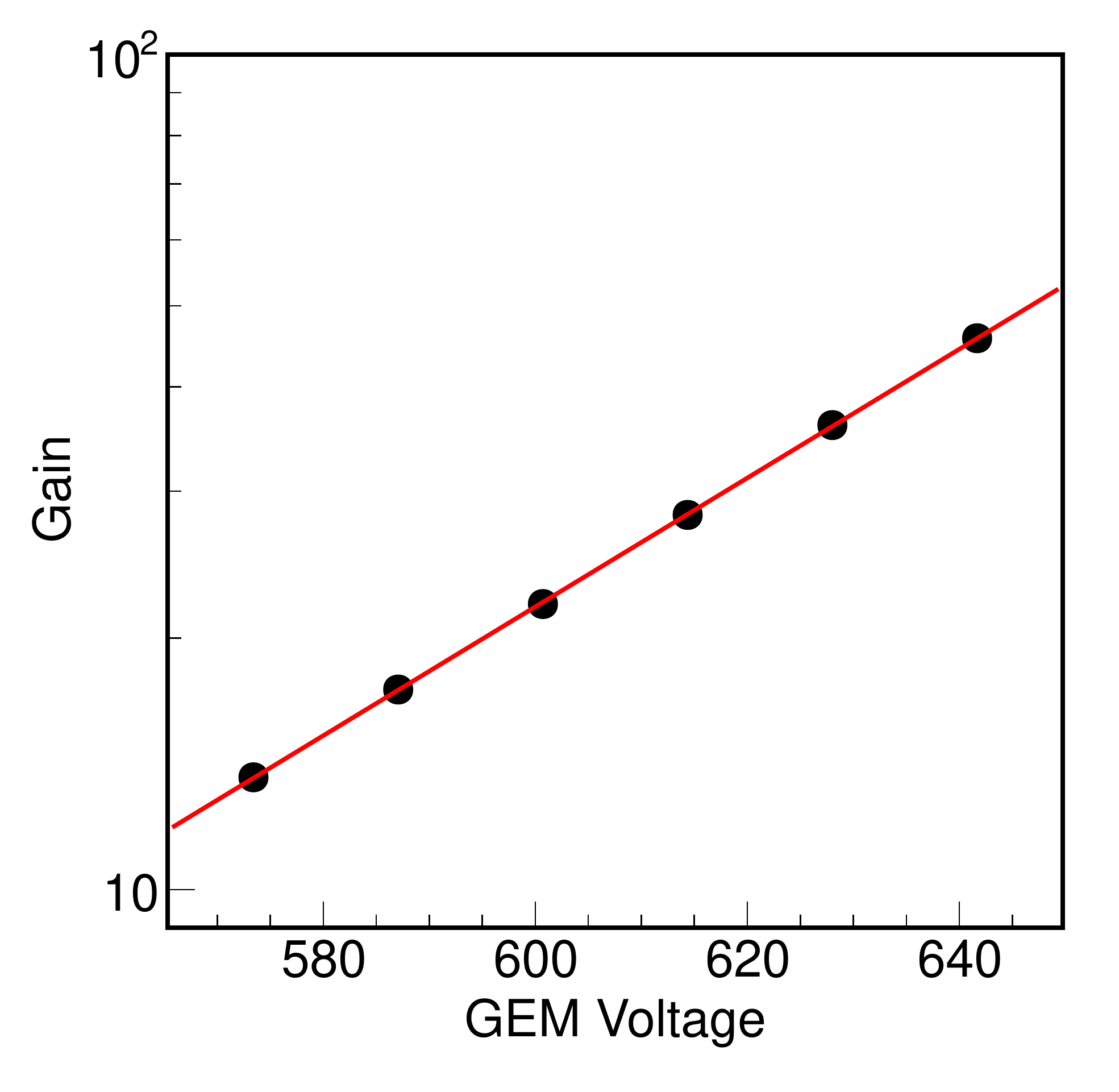}
\setlength{\abovecaptionskip}{0pt}
\caption{Effective gain of double GEM layer versus sum of GEM voltages. Top: measured for high GEM voltages with 5.9-keV x-rays in ArCO$_2$ gas (black points) and HeCO$_2$ gas (gray points). Bottom: measured for lower GEM voltages with 4-MeV alpha particles in ArCO$_2$ gas. The straight lines show fits of equation \ref{eq:gain} to the data.\label{fig:gain}}
\setlength{\belowcaptionskip}{0pt}
\end{figure}

\begin{figure}
\centering
\includegraphics[width=8cm]{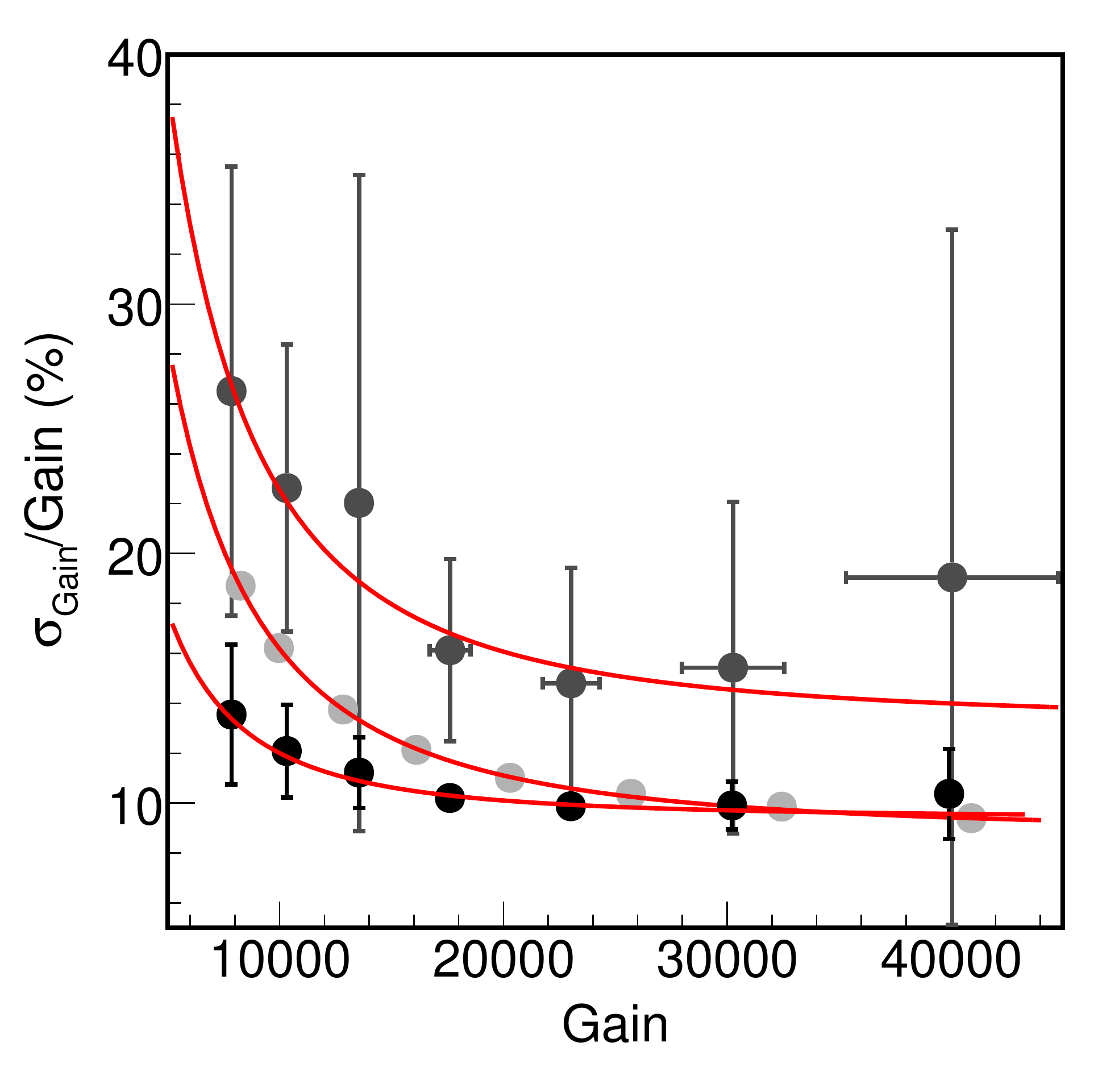}\\\includegraphics[width=8cm]{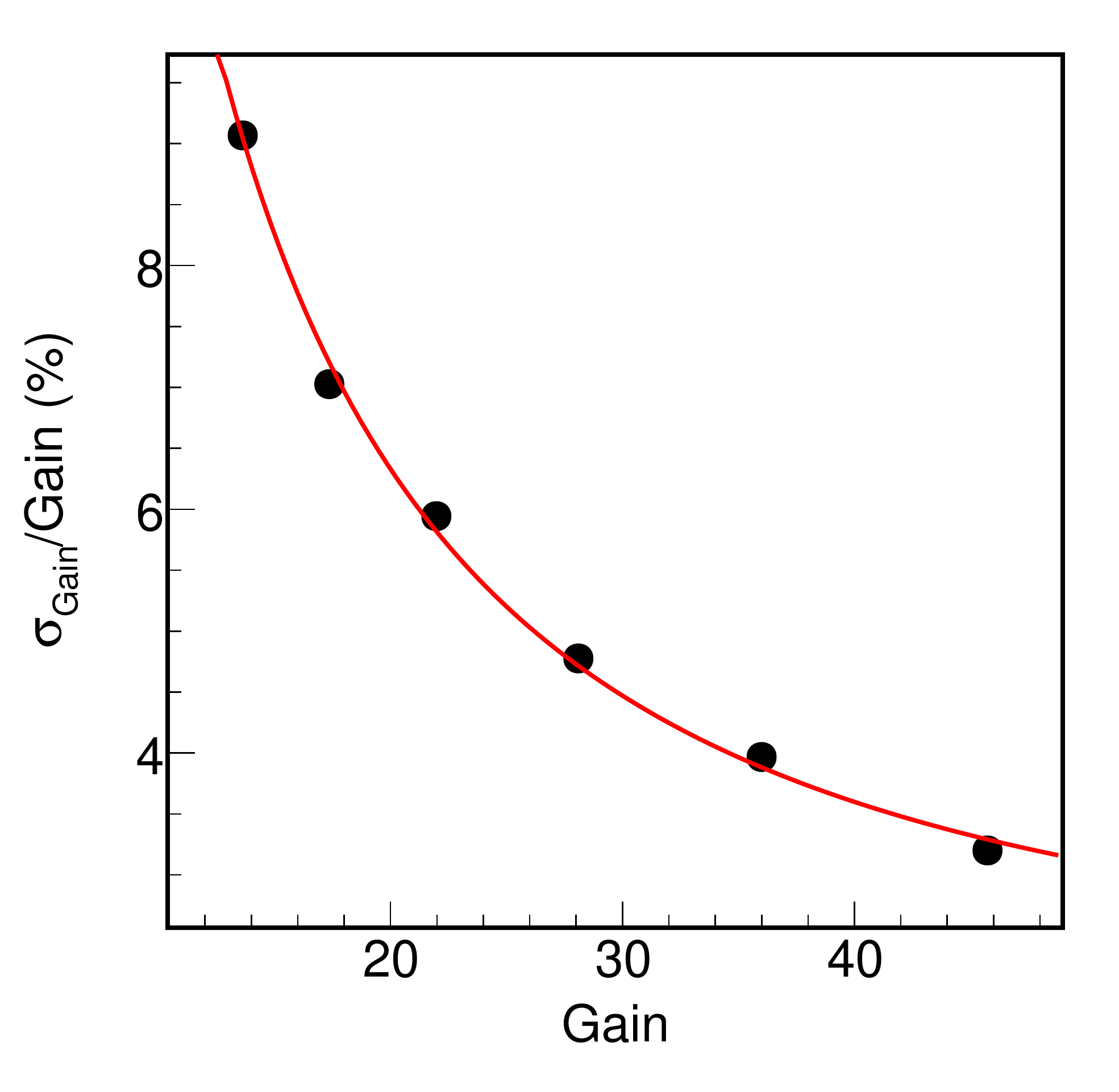}
\setlength{\abovecaptionskip}{0pt}
\caption{Gain resolution versus effective gain of double GEM. Top: measured for high GEM voltages with Fe-55 x-rays in ArCO$_2$ gas (black points: 5.9~keV, dark gray points: 2.9~keV) and HeCO$_2$ gas (light gray points, 5.9~keV). Bottom: measured for low GEM voltages with Po-210 alpha particles in HeCO$_2$ gas. The curves show fits of equation \ref{eq:resvsgain} to the data.\label{fig:gain_resolution}}
\setlength{\belowcaptionskip}{0pt}
\end{figure}

\subsection{Dependence on GEM Voltage, Gas Type, and Energy}
We were able to operate the detector stably and for weeks at a time with double GEM gains as high as $4\times 10^4$. Figure~\ref{fig:gain} summarizes the measured gain versus double-GEM voltage, where each gain value was obtained from a fit to a PHA spectrum as discussed above. The dependence of gain on GEM voltage is well described by the function
\begin{equation}
\begin{aligned}
G &= 10^{\frac{V_{\rm GEM}-V_1}{V_2}},
\label{eq:gain}
\end{aligned}
\end{equation}
where $G$ is the gain, $V_{\rm GEM}$ is the sum of the voltages across the two GEMs, and $V_1$ and $V_2$ are free parameters that we extract with a $\chi^2$ fit, see Table \ref{table:gain}. These parameters agree at the 20\%-level with measurements of our previous prototype \cite{Kim:2008zzi} and measurements by other groups \cite{Bachmann:1999xc}. When extrapolated to the same GEM voltages, the effective gain measured with the Po-210 alpha source is slightly lower than that measured with the Fe-55 x-ray source. This may be due to an increased loss of ionization to recombination in the case of alpha particles. 

Figure~\ref{fig:gain_resolution} summarizes the measured gain resolution versus effective gain, where each resolution and gain value was obtained from a fit to a PHA spectrum as discussed above. The resolution versus gain is well described by a function of form
\begin{equation}
\begin{aligned}
 \sigma_{G}/G &=\sqrt{(a/G)^2+b^2}, \label{eq:resvsgain}
\end{aligned}
\end{equation}where G is the gain, {\it a} a term due to noise fluctuations, and {\it b} is the asymptotic detector resolution at high gain. Fitting this function to the data yields the parameters shown in Table \ref{table:gain}. The gain resolution at 5.9 keV becomes asymptotic (at high gain, where PHA noise becomes negligible) to $\approx 9\%$, typical for gas detectors. The measured resolution for MeV signals is as low as 3.4\%, with the fitted function extrapolating to an asymptotic value of ~2\% at higher gain, i.e. approaching the excellent energy resolution of solid state detectors.

\begin{figure}
\centering
\includegraphics[width=8cm]{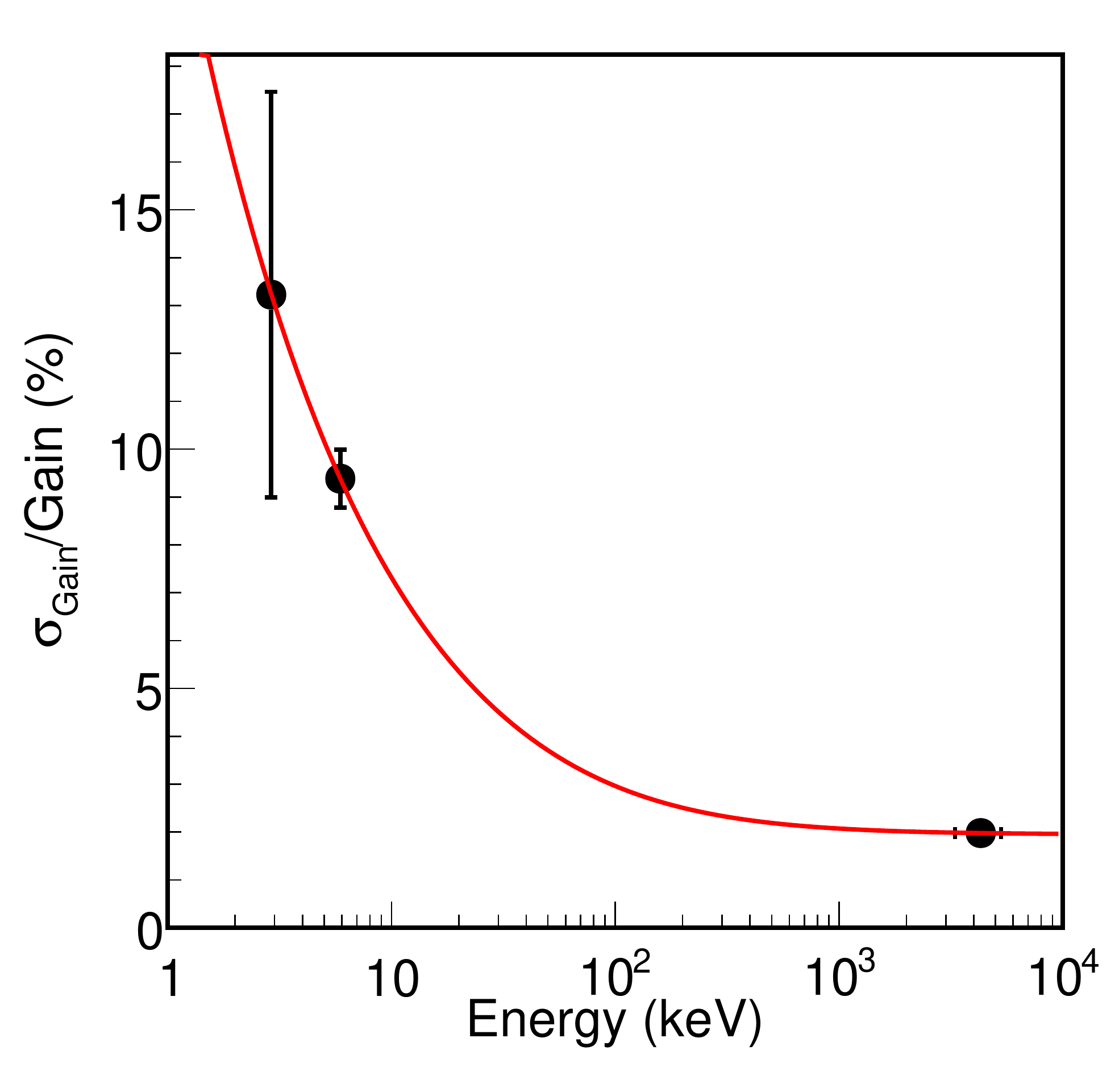}
\setlength{\abovecaptionskip}{0pt}
\caption{Asymptotic (high-gain) gain resolution versus ionization energy for ArCO$_2$ gas. See text for discussion. \label{fig:resolution_vs_energy}}
\setlength{\belowcaptionskip}{0pt}
\end{figure}

\begin{table*}
\begin{center}
	\begin{tabular}{|l|l|l|l|l|l|}
 	\hline    
    gas mixture	&  ionization energy	& $V_1$ 	& $V_2$	& a		& b  \\ \hline
    ArCO$_2$	& 5.9~keV      	& $434\pm9$    	& $116\pm2$  	&   $(7.8\pm2.2)\times 10^4$  	&  $(9.4\pm0.6)$\%           \\ 
    ArCO$_2$	& 2.9~keV      	& $434\pm18$	& $116\pm4$ 		&   $(1.8\pm0.7)\times 10^5$  	&  $(13\pm4.2)$\%           \\ 
    ArCO$_2$  	&  4.0~MeV      	& $427\pm3$		& $130\pm2$    	&   $120.3\pm0.7$ 	&  $(1.98\pm0.06$)\%              \\  
    HeCO$_2$	&  5.9~keV      	& $356\pm7$ 		& $136\pm2$   	&  $(1.375\pm0.003)\times 10^5$  	&  $(8.79\pm.02)$\%               \\  \hline
    \end{tabular}
\caption{Parameters characterizing the double GEM gain and its resolution, defined by equations \ref{eq:gain} and \ref{eq:resvsgain}, and with the following physical interpretations: $V_1$ is the double-GEM voltage that gives a gain of unity. $V_2$ is the double-GEM voltage change required to increase the gain by a factor of 10. The parameter {\it a} is the gain above which the gain resolution becomes asymptotic, and {\it b} is the asymptotic gain resolution at high gain.
 \label{table:gain}}
\end{center}
\end{table*}

Since three of the resolution measurements were performed in ArCO$_2$, but at different energies, we can obtain the dependence of the asymptotic (high-gain) detector resolution versus energy, as shown in Fig.~\ref{fig:resolution_vs_energy}. The three points are well described by the function 
\begin{equation}
\begin{aligned}
\sigma_{G}/G=\sqrt{d^2+c^2/E}, 
\end{aligned}
\end{equation}
with $d=(1.94\pm0.07)\%$, and $c=(22.3\pm1.5)\%\times\sqrt {keV}$, where G is the gain, {\it d} is the effective gain stability of the detector and the measurement system (e.g. limited by the stability of the GEM high voltage supply, shaping amplifier, and PHA), while $c/\sqrt E$ is the fundamental gain resolution of the technique, determined by statistical fluctuations in the initial ionization statistics and in the avalanche process. These two statistical effects both average out as the number of primary electrons increases, which gives rise to the factor $1/\sqrt E$ multiplying the $c$ term. It can be shown \cite{Knoll:2000fj} that $c^2=W\times (F+b')$, where $W$ is the work function of the gas, $F$ is the Fano factor describing the fluctuations in primary ionization statistics \cite{Fano:1947zz}, and $b'$ is the Polya distribution parameter describing the variation in the avalanche gain \cite{polya}. Using our measured value for $c$, $W=28.05~eV/{\rm electron}$ and the Fano factor $F=0.23$ for gaseous Argon \cite{Fano2}, we obtain $b'=1.54$ for the combined double GEM layer, or $b=0.77$ per GEM.       

Figure~\ref{fig:resolution_vs_energy} leads to interesting conclusions: For the lowest energy measured, 2.9 keV, the gain resolution is of order 15\%, likely sufficient to provide head/tail discrimination for keV-scale nuclear recoils, such as might be expected for low mass ($\approx$ 10-GeV) WIMPs. Since the resolution in the keV range is limited by the primary ionization statistics in the gas and the avalanche gain fluctuations in the GEMs, improving the detector electronics would not improve energy resolution in this energy region. Electron counting \cite{Sorensen:2012qc}, however, may reduce or theoretically even fully eliminate the component of energy resolution due to gain fluctuations, and thus lead to improved energy resolution in the keV regime. At MeV scale energies, of interest for fast-neutron spectroscopy, the number of primary electrons is so large that statistical fluctuations in ionization statistics and avalanche gain become lesser effects, and instead the detector and measurement system stability currently limit the gain resolution measured. Hence improved pixel, shaper, and PHA electronics could lead to improved gain and energy resolution in this regime.
 
\section{Measurements of Point Resolution\label{sec:pointres}}

\begin{figure}
\centering
\includegraphics[width=9cm]{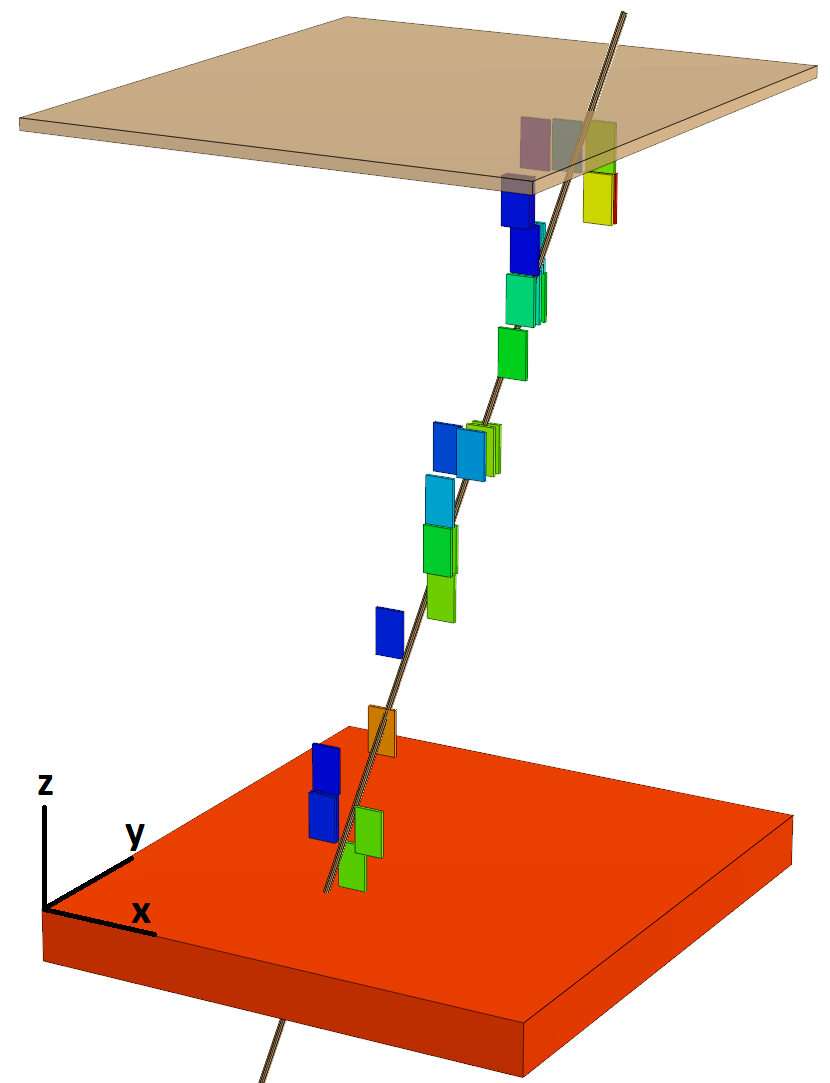}
\smallskip
\includegraphics[width=9cm]{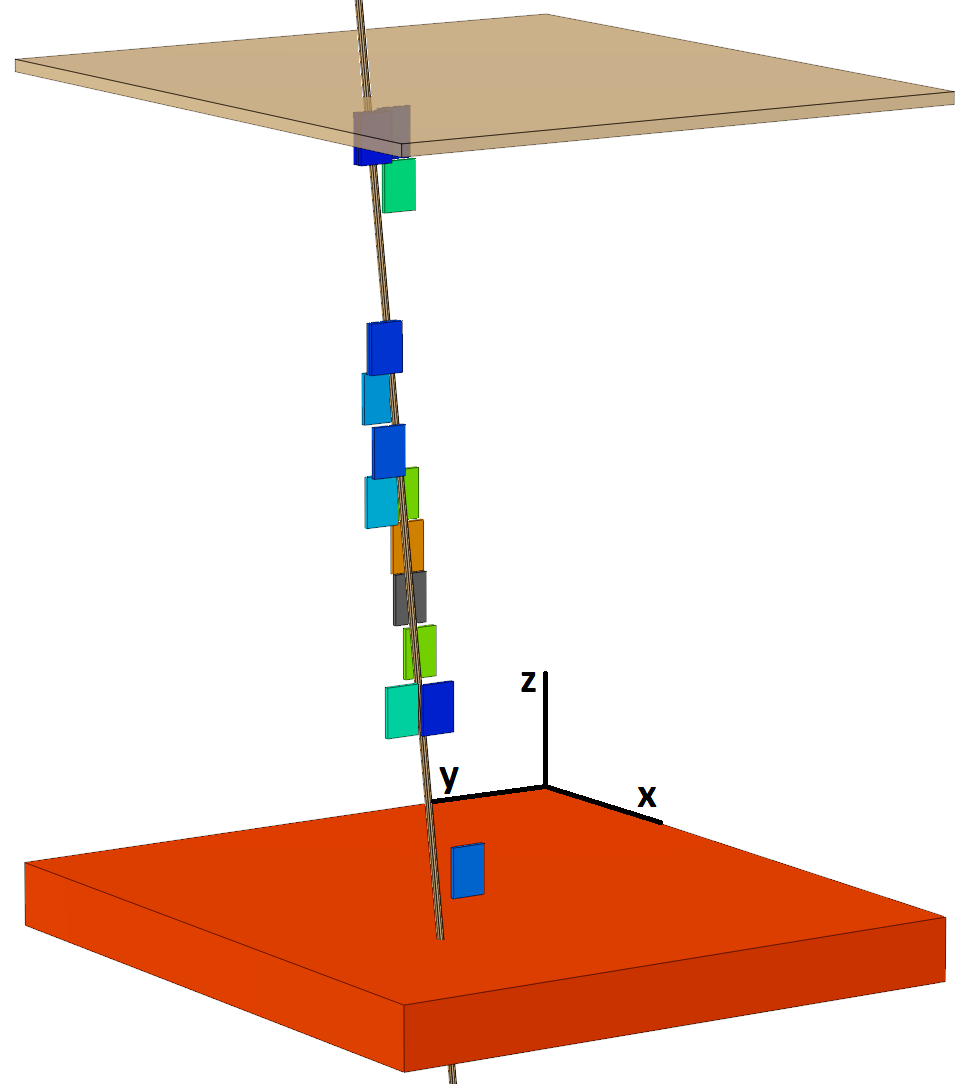}

\setlength{\abovecaptionskip}{0pt}
\caption{Cosmic ray events measured in ArCO$_2$ gas (top) and HeCO$_2$ gas (bottom) with a detector gain of $4\times 10^4$. The small boxes represent hits recorded by the pixel chip, and measure $50~\mu{\rm m}~\times~400~\mu{\rm m}$ in $x \times y$. Their color (available online) is determined by the measured ToT, which reflects the ionization density. The line is the best fit to the hits. The large volume below the hits is the pixel chip, and the transparent volume above the hits is part of the cathode mesh. Assuming these tracks were created by minimum ionizing particles, the total energies deposited are of order 2~keV (ArCO$_2$) and 1.0~keV (HeCO$_2$), demonstrating the excellent detector sensitivity at high gain settings.
\label{fig:cosmic}}
\setlength{\belowcaptionskip}{0pt}
\end{figure}

\begin{figure}
\centering
\includegraphics[width=9cm]{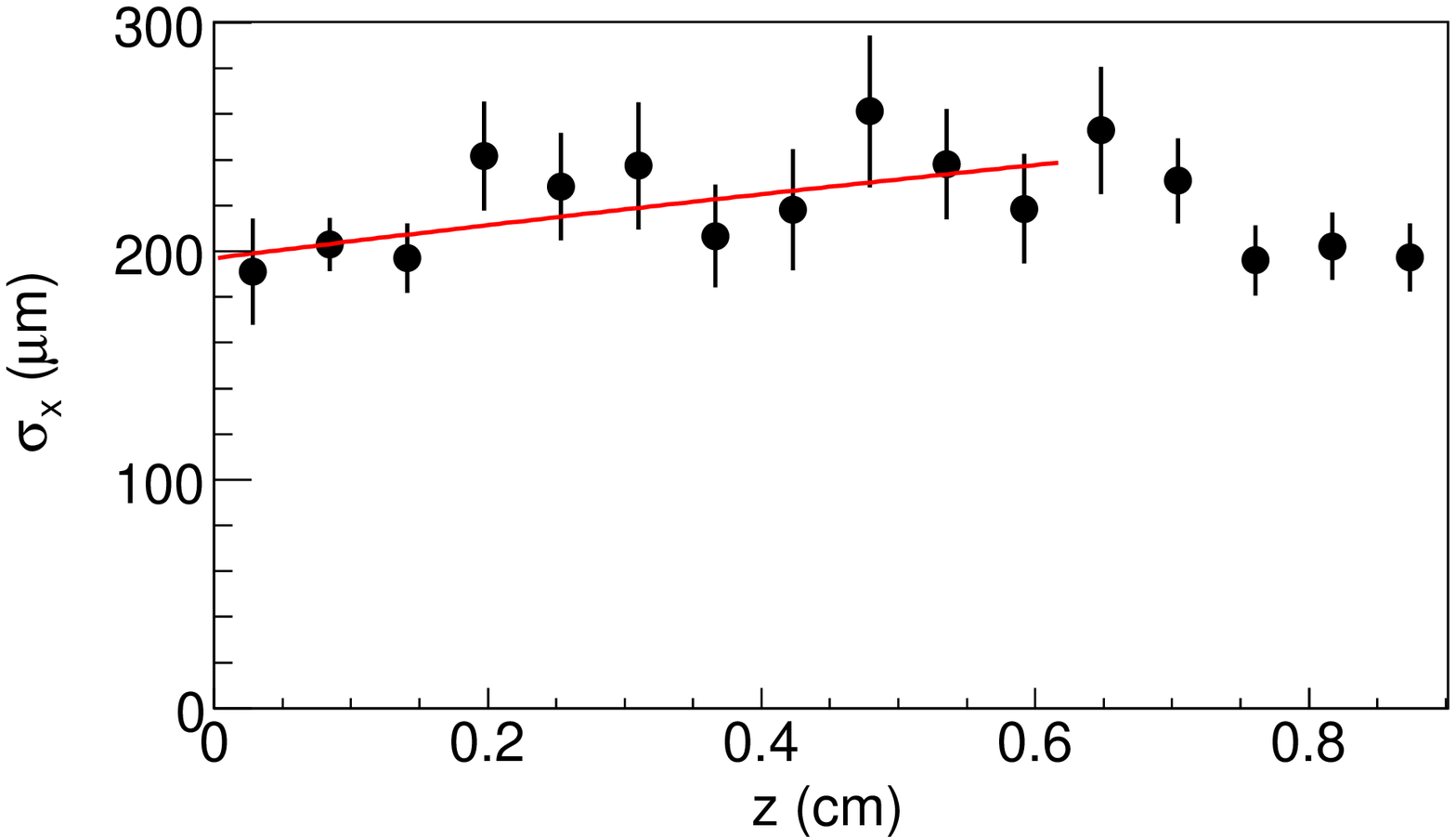}
\includegraphics[width=9cm]{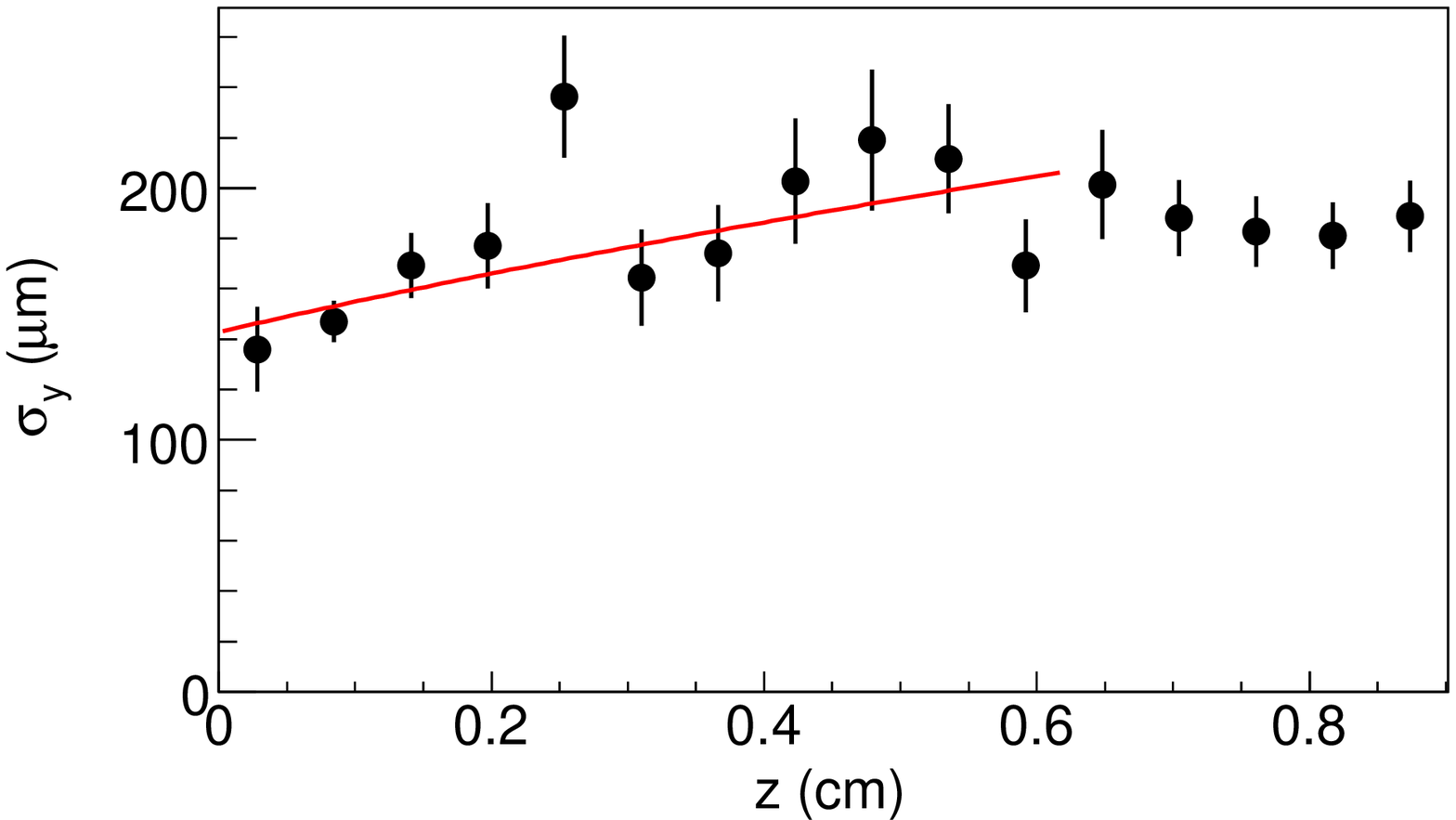}
\setlength{\abovecaptionskip}{0pt}
\caption{Measured transverse point-resolution in $x$ (upper) and $y$ (lower) versus drift length ($z$) for cosmic ray tracks in HeCO$_2$ gas. The black points show experimental data, the curves (red in online version) are fits of equation \ref{eq:res} to the data.}
\label{fig:point_resolution}
\setlength{\belowcaptionskip}{0pt}
\end{figure}

We use cosmic ray events to measure the transverse ($x,y$) point resolution of the detector with HeCO$_2$ gas, as a function of the drift length, $z$.  For this study we operate the detector with high gain ($>40\times 10^3$) and increased drift field (1180~V/cm) for an expected drift velocity of 22.8~${\rm \mu m/ns}$ \cite{magboltz, Biagi:1999nwa}. Since the chip is sampling charge at a fixed 40~MHz, the increased drift velocity degrades the $z$-resolution, but allows 16 pixel chip time bins, the maximum number that can be recorded consecutively, to cover the entire drift gap. This enables an important analysis technique: we select tracks that traverse the entire drift gap and which have hits near the top and bottom of this gap, as shown in Fig.~\ref{fig:cosmic}. This allows us to assign an absolute $z$ position to each recorded hit, which is normally not possible in a TPC. For each event, we fit a straight line to the distribution of hits and determine the $x$ and $y$ point resolution from the distribution of the $x$-distance and $y$-distance between the hit positions and points of closest approach to the line. Though these distances are not generally equal to the true miss-measurements, the two quantities become identical for vertical tracks. We keep tracks within 25 degrees of vertical ($z$). A GEANT4\cite{Agostinelli:2002hh} Monte Carlo simulation estimates that for such tracks our procedure measures the actual $x$ and $y$ point resolutions to an accuracy better than 5 percent. We perform this procedure for each of the 16 $z$-coordinates separately, and obtain the transverse point resolution versus drift distance $z$, shown in Fig.~\ref{fig:point_resolution}. We expect the point resolution to be the quadrature sum of the readout plane's point resolution (which is independent of $z$), and the transverse diffusion (which is proportional to $\sqrt z$), i.e. 
\begin{equation}
\begin{aligned}
 \sigma^T_{x/y}(z)=\sqrt{(\sigma^{R}_{x/y})^2+C_T^2 z},
\end{aligned}
\label{eq:res}
\end{equation}
where $\sigma^{R}_{x/y}$ is the readout plane point resolution in $x$ or $y$, $C_T$ is the transverse diffusion per $\sqrt{z}$ , and $z$ is the drift distance.  By fitting this function to the experimental data, we obtain $\sigma^{R}_x=(197 \pm 11)~\mu m$ and $\sigma^{R}_y=(142 \pm 9)~\mu m$, in good agreement with the analytical estimates $\sigma^{R}_x=184~\mu m$ and $\sigma^{R}_y=143~\mu m$. Table \ref{table:pointresolution} gives a breakdown of the analytical estimate. The resolution in $y$ is better than that in $x$ because the rectangular pixels are smaller in the $y$ direction ($50~\mu m$) than the $x$-direction ($400~\mu m$).  Note that in $y$, the readout resolution is not limited by the feature size of either the GEMs or pixels, but by the diffusion in the collection and transfer gaps. 
\begin{table}
\begin{center}
	\begin{tabular}{|l|l|l|l|}
 	\hline    
    							& $\sigma_x (\mu m)$ & $\sigma_y (\mu m)$ \\ \hline
    GEM1 hole spacing  	& 40.4      		& 40.4	      \\ 
    transverse diffusion in collection gap 			& 93.2	      	& 93.2		   \\ 
    GEM2 hole spacing		& 40.4      		& 40.4	        \\ 
    transverse diffusion in transfer gap		 	& 91.8    		& 91.8       	\\ 
    pixel size		 		& 115	      	& 14.4	        \\ \hline
    Predicted $\sigma_{x,y}$ &	 184  	&  143  	       \\ 
    Measured $\sigma_{x,y}$ & 	 $197 \pm 11$    	&   $142 \pm 9$	      \\ \hline
    \end{tabular}
\caption{Estimated contributions to the readout plane resolution. The collection gap, transfer gap, and coordinate system are defined in Fig.~\ref{fig:detectordrawing}. Diffusion values were calculated with Magboltz  \cite{magboltz} and the GEM and pixel resolution values are the respective feature sizes divided by $\sqrt{12}$. The predicted resolution is the quadrature-sum of the contributions. 
 \label{table:pointresolution}}
\end{center}
\end{table}
As for the diffusion in the drift gap, the fits to $\sigma_{x}(z)$ and $\sigma_{y}(z)$ yield $C_T=(172\pm 46)~\mu m$ and $C_T=(189\pm 27)~\mu m$, respectively, where the quoted uncertainties are statistical only. The two values are consistent with each other, as expected, but significantly larger than the Magboltz prediction of $C_T=109~\mu {\rm m}$. The discrepancy may be due to non-uniformities in the drift field, or possibly from impurities in the gas. However, the GEANT4 simulation also suggests that while our analysis procedure is reliable for measuring the point resolution, it adds a systematic uncertainty to the measured diffusion: this is because equation \ref{eq:res} does not account for detector inefficiencies in the case of highly diffuse charge at the edge of the ionization cloud, which can be seen to bias the measured diffusion in Fig. \ref{fig:point_resolution} for $z$ greater than 0.6~cm. Hence we exclude this $z$-range from the fit. However, the exact diffusion value measured is very sensitive to the exact $z$-range used, resulting in a systematic uncertainty of order 50\%. If accurate diffusion measurements are desired in the future, this may be accomplished by operating the detector at settings where the single electron efficiency approaches unity, so that the fraction of undetected charge becomes negligible.

\section{3-D Tracking of Helium Nuclei: Angular Resolution\label{angles}}

\begin{figure}
\centering
\includegraphics[width=8.8cm]{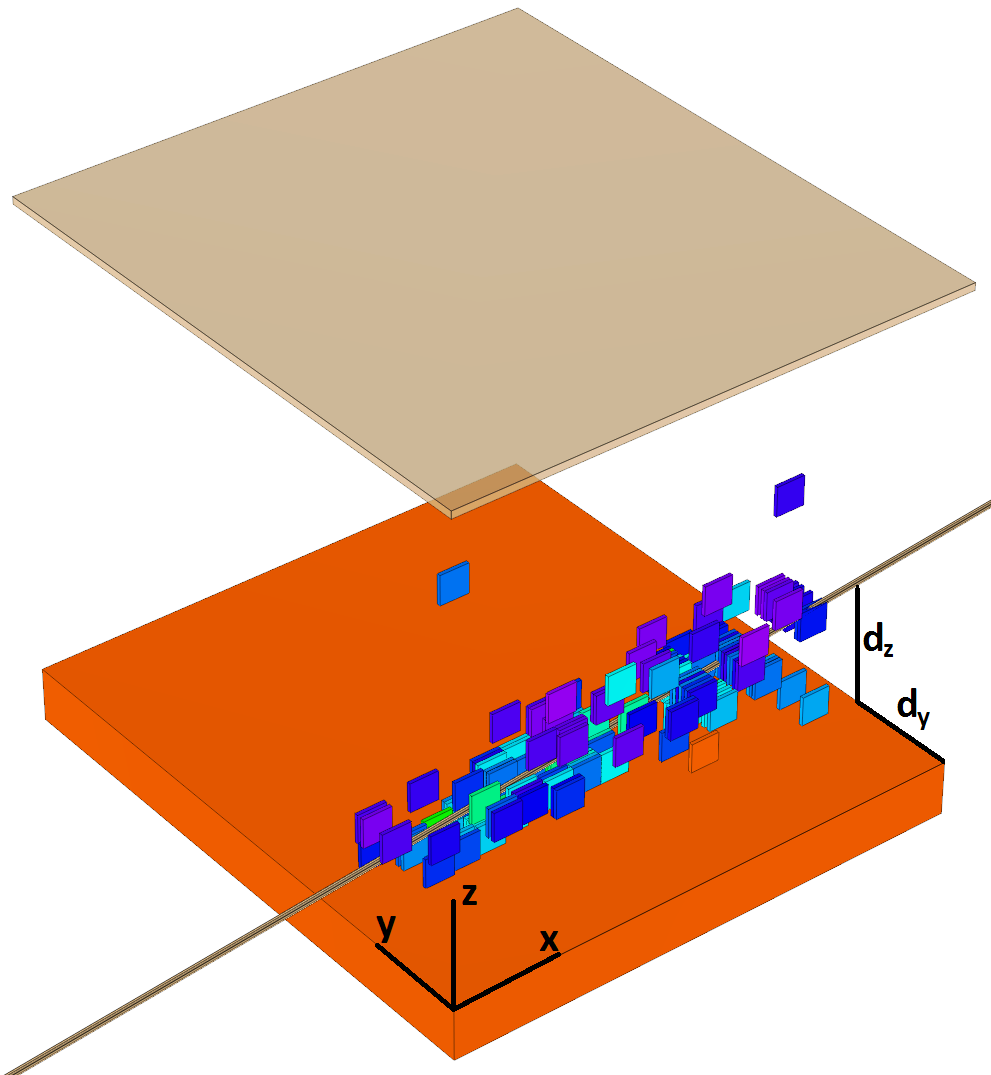}
\setlength{\abovecaptionskip}{0pt}
\caption{Alpha particle track segment measured in HeCO$_2$ gas with a detector gain of $3\times 10^3$. The Po-210 alpha source is located beyond the right edge of the image, at $x\approx 46~{\rm mm}$. \label{fig:alpha}}
\setlength{\belowcaptionskip}{0pt}
\end{figure}

One incidental benefit of using $^4{\rm He}$ nuclei as our neutron target is that we can use alpha particle sources to estimate the detector performance for detecting He-recoils. We estimate the angular resolution with three complementary methods, described in the next three paragraphs.
\begin{figure}
\centering
\includegraphics[width=8cm]{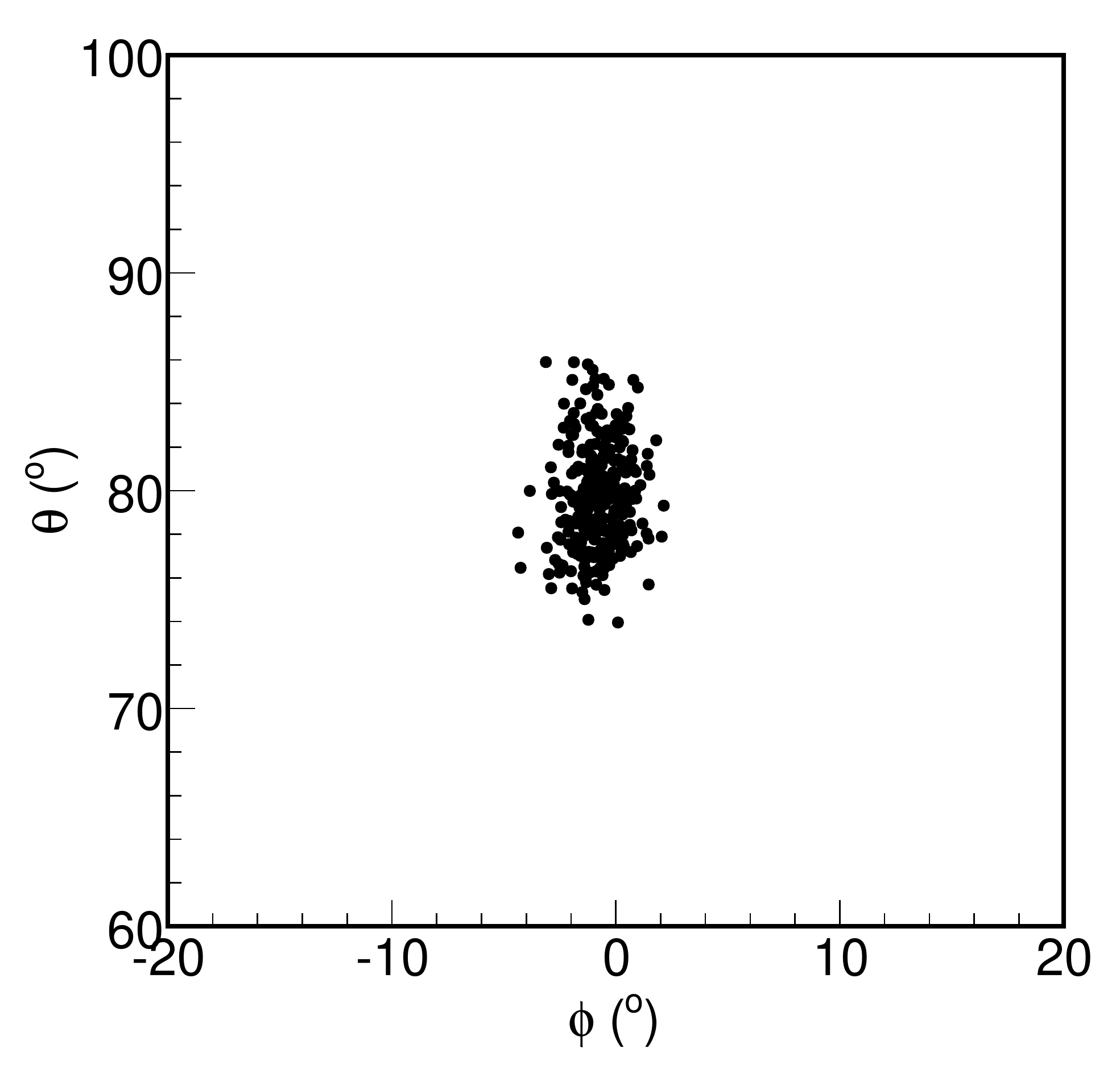}\\
\includegraphics[width=8cm]{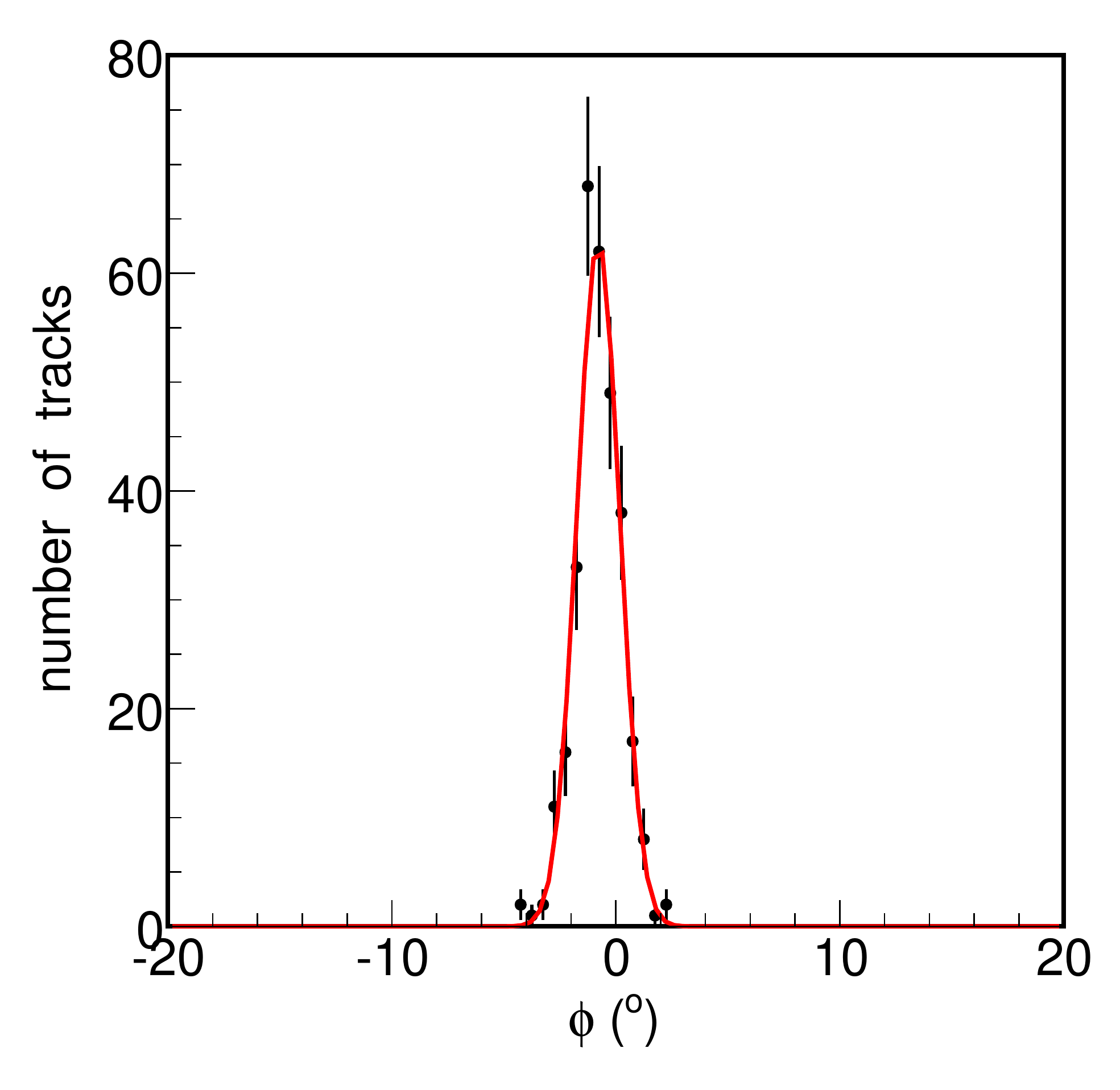}
\includegraphics[width=8cm]{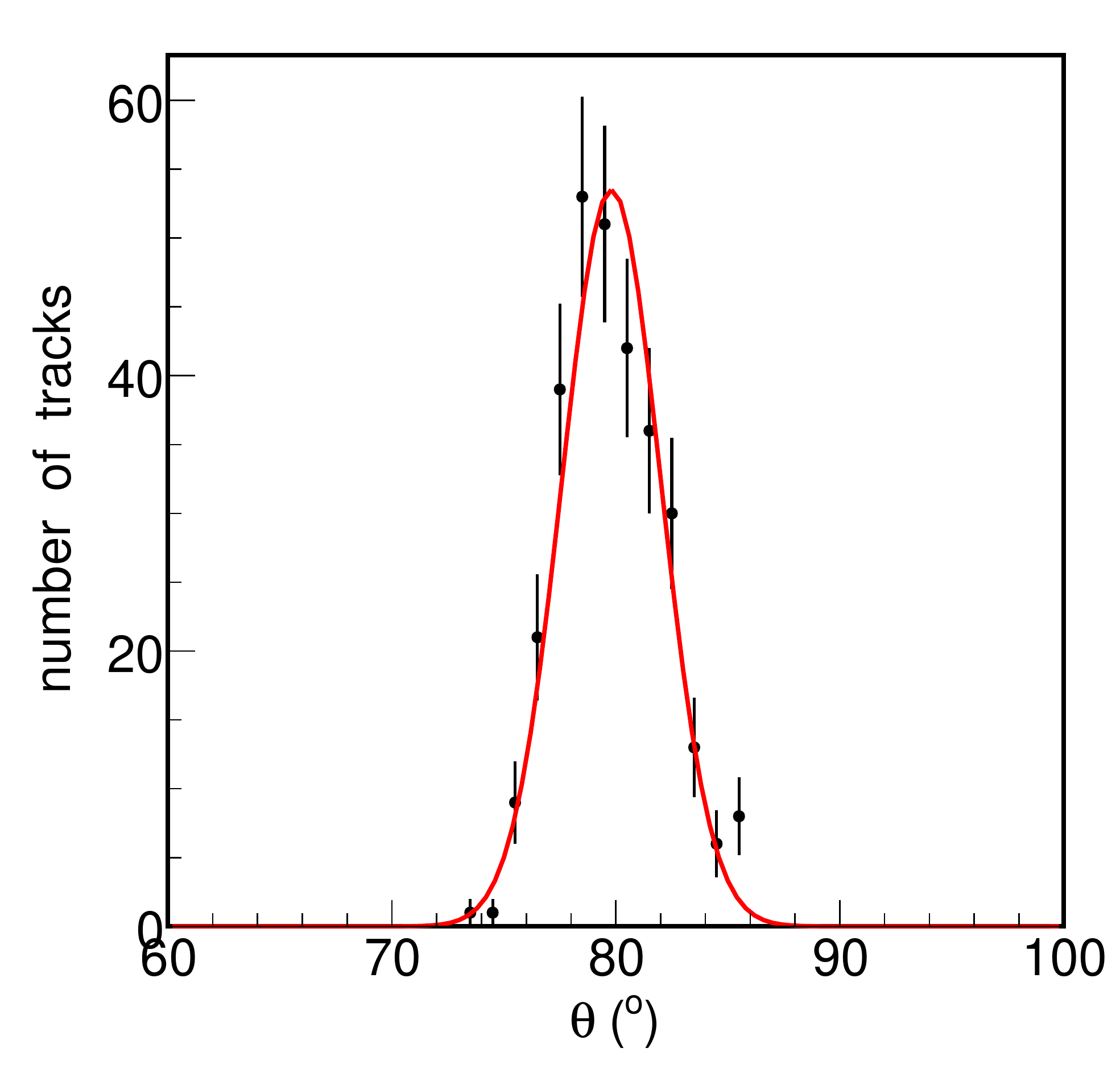}
\setlength{\abovecaptionskip}{0pt}
\caption{Reconstructed direction of Po-210 alpha tracks in HeCO$_2$ gas, with the alpha source placed 3.88~cm from the detector. Top: polar angle ($\theta$) versus azimuthal angle ($\phi$) for each reconstructed track. Middle: azimuthal angle ($\phi$) distribution (black data points) with a Gaussian fit superimposed. Bottom: polar angle ($\theta$) distribution (black data points) with a Gaussian fit superimposed.}  
\label{fig:alpha_theta_phi}
\setlength{\belowcaptionskip}{0pt}
\end{figure}
All three methods make use of the same data sample, recorded with a Po-210 (5.3 MeV) alpha source located at $z\approx 5~{\rm mm}$, $x\approx 46~{\rm mm}$. The source is oriented so that the emitted alpha particles traverse the drift gap above the chip, approximately parallel to the $x$-axis. Due to the large ionization density in these events, we operate the detector with a reduced gain of 3200. We select a pure sample of well-reconstructed alpha tracks by fitting the events with a straight line and requiring $L>5000~\mu {\rm m}$, $\chi^2/n.d.f. < 2.0$, and $N>60$, where $L$ is the track length, $\chi^2/n.d.f.$ is the reduced chi-squared, and $N$ is the number of pixels hit. To ensure the full energy of tracks can be measured, we veto events with hits within $400~\mu {\rm m}$ of the $y$-edges of the chip. Figure~\ref{fig:alpha} shows an example of an event passing the selection. 

Method I sets an upper limit on the track angle resolution using the angular spread of tracks pointing back to the alpha source. Though the physical source is not collimated, we collimate it in the analysis by requiring $3.0~{\rm mm}<d_y<3.4~{\rm mm}$, where $d_y$ is the $y$-impact parameter of the fitted track at $x=7.2~{\rm mm}$, the chip edge closest to the alpha-source. Figure~\ref{fig:alpha_theta_phi} shows the polar and azimuthal angle distribution for the 3-D tracks passing that selection. These tracks point back to a point-like object, as expected for the small disk source used. By fitting a Gaussian distribution to the 1-D projections in Fig. \ref{fig:alpha_theta_phi}, we obtain $\sigma^{source}_\phi=(0.95\pm 0.04) ^\circ$, and $\sigma^{source}_\theta=(2.2\pm 0.1)^\circ$.  These values are upper limits on the track angle resolutions, because $\sigma^{source}$ is the convolution of several effects:

\begin{equation}
\begin{aligned}
 \sigma^{source} = \sqrt{\sigma^2_{\rm DET}+\sigma^2_{\rm STRAG}+\sigma^2_{\rm SIZE}+\sigma^2_{\rm COLL}},
\end{aligned}
\label{eq:ang_res}
\end{equation}
where $\sigma_{DET}$ is the true track angle resolution of the detector, and the remaining terms represent broadening of the observed angular distribution due to nuclear straggling (kinked tracks), the size of the alpha-particle source, and the lack of source collimation, respectively. Without a selection on impact parameter $\sigma_{\rm COLL}$ is dominant for our geometry. In the $\phi$ direction, the selection on $d_y$ ensures that the contribution from collimation is sub-dominant ($\sigma_{\rm COLL}\lesssim 0.3^\circ$), so that $\sigma_{\rm SIZE}$ ($\approx 1^\circ$) dominates.  Since, however, a TPC cannot measure absolute position in the drift direction ($z$), we cannot know the true $d_z$, and hence cannot select on it to collimate the source in theta. This is the reason why $\sigma^{source}_\theta > \sigma^{source}_\phi.$ We have not tried to experimentally measure the straggling.

\begin{figure}
\centering
\includegraphics[width=8cm]{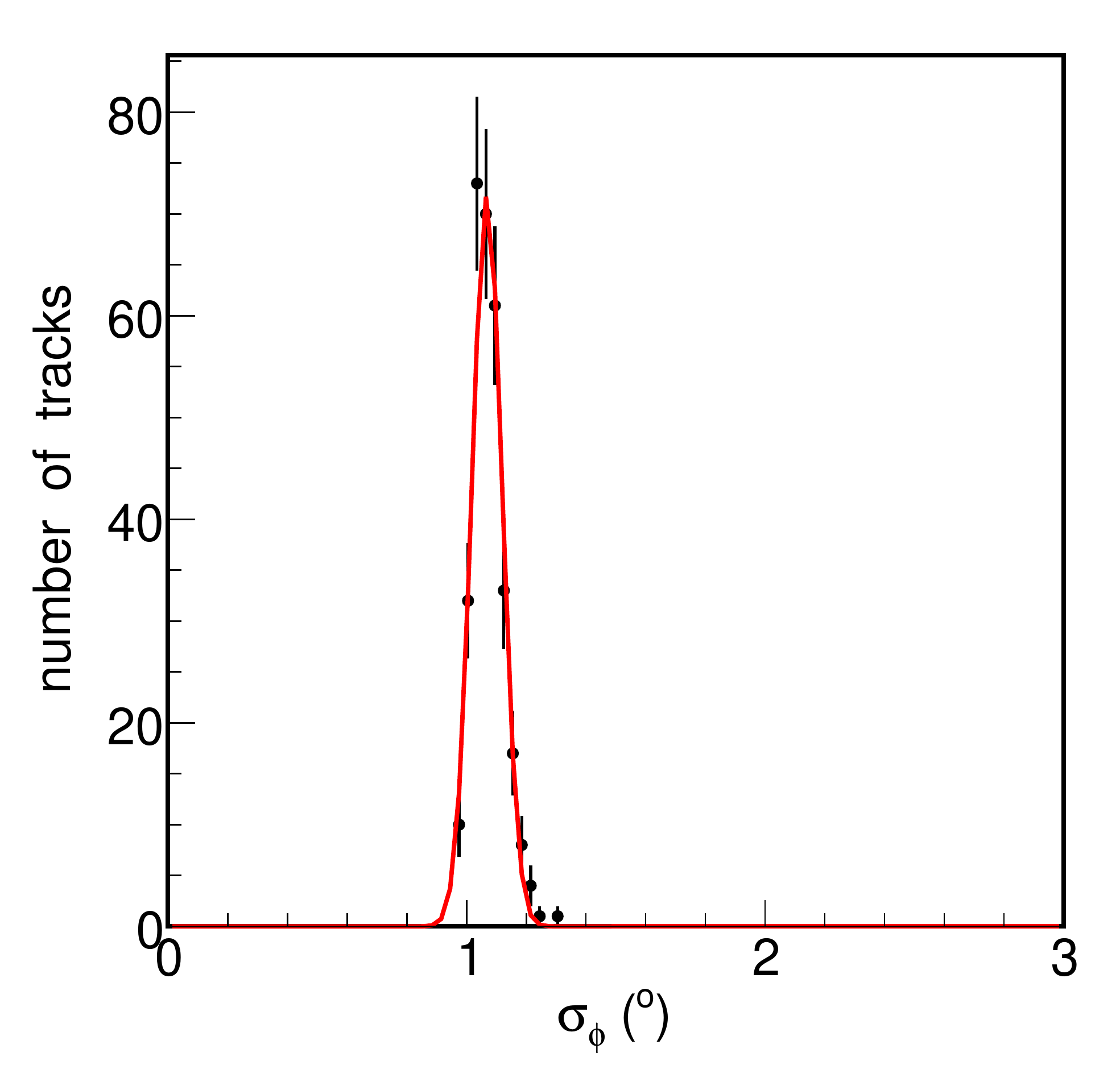}
\includegraphics[width=8cm]{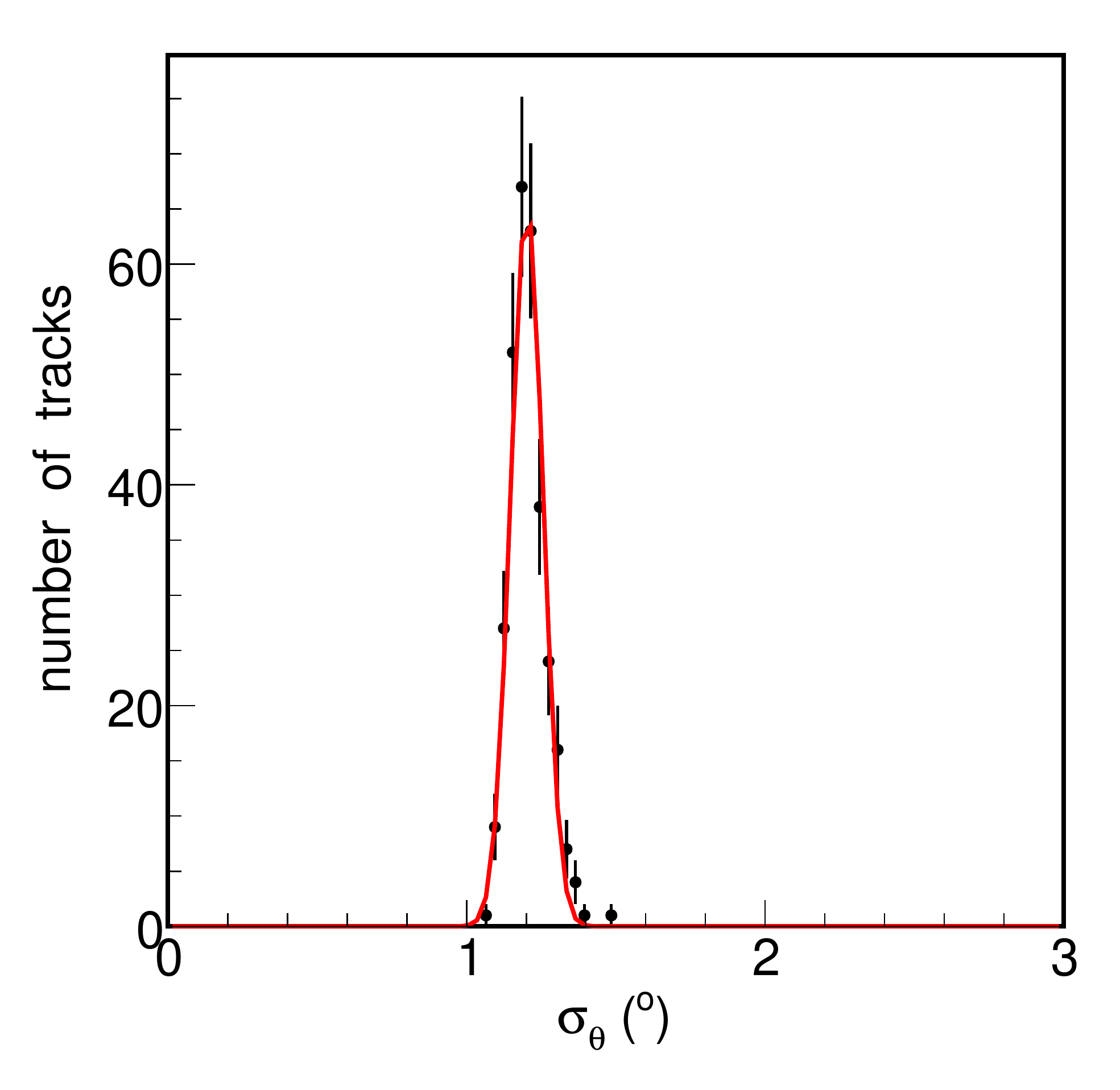}
\setlength{\abovecaptionskip}{0pt}
\caption{Track fitter uncertainties (black data points) in azimuthal angle ($\phi$) (top) and polar angle ($\theta$) (bottom), for alpha particle events measured in HeCO$_2$ gas. The curves show Gaussian fits to the data. \label{fig:fitter_errors}}
\setlength{\belowcaptionskip}{0pt}
\end{figure}

Method II estimates the angular resolution from the event-by-event uncertainties reported by the track-fitter. This requires that the $\chi^2$ used to fit the tracks is calculated with the correct point resolutions. We use the residual distributions from the alpha track sample to obtain $\sigma_y=(275\pm 0.7)~\mu {\rm m}$, and $\sigma_z=(313\pm 62)~\mu {\rm m}$. Because most alpha tracks are nearly parallel to the $x$-axis, these events don't really constrain $\sigma_x$. Since, except for the pixel size, our detector is symmetric in $x$ and $y$, we instead obtain $\sigma_x=(298\pm1)~\mu {\rm m}$ by subtracting the pixel y-resolution (Table \ref{table:pointresolution}) in quadrature from $\sigma_y$, and then adding the pixel x-resolution in quadrature. Our final results are not sensitive to the exact choice of $\sigma_x$. Note that these point resolutions for alpha particles are slightly larger than the point resolutions measured for cosmic ray events (Fig. \ref{fig:point_resolution}) at $z=0.5$~cm (the approximate vertical position of the alpha source). This, presumably, is mainly due to straggling. The track angle resolutions reported by the track fitter are shown in Fig. \ref{fig:fitter_errors}. Fitting for the mean resolution with a Gaussian gives $\sigma_{\phi}=(1.068\pm0.003)^\circ$ and $\sigma_{\theta}=(1.202\pm0.004)^\circ$. These values are again upper limits on the track angle resolutions, because we estimated the point resolution without removing outlier hits or accounting for kinked tracks. Compared with Method I, the present technique is insensitive to the direction in space where each track points, but rather measures how well the hits on each track align in a given direction. As a result, the present technique excludes the contributions from $\sigma^2_{\rm COLL}$ and $\sigma^2_{\rm SIZE}$, which dominated in Method I. This explains why Method II gives a stronger limit on $\sigma_{\theta}$. 
\begin{figure}
\centering
\includegraphics[width=9cm]{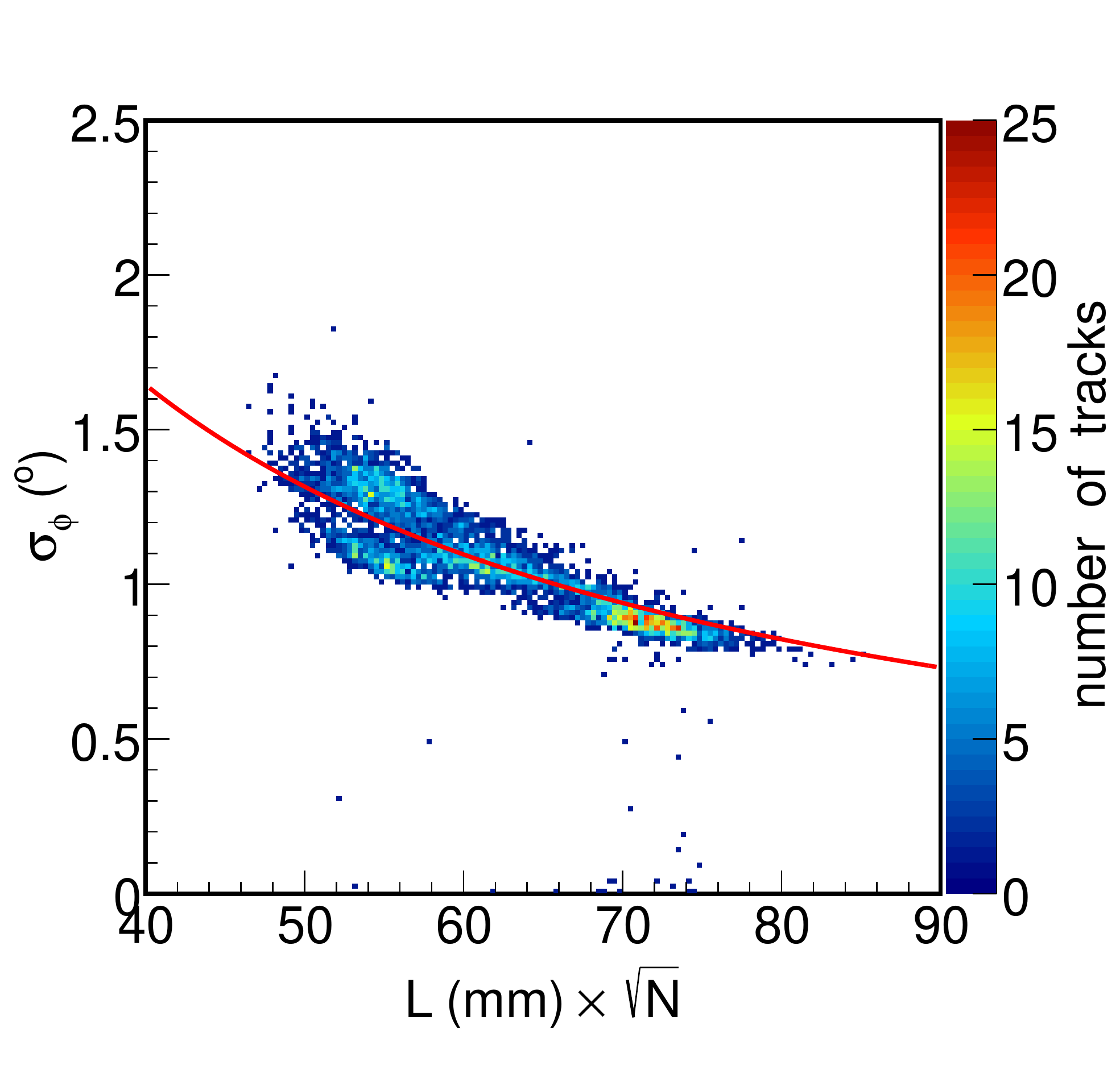}
\includegraphics[width=9cm]{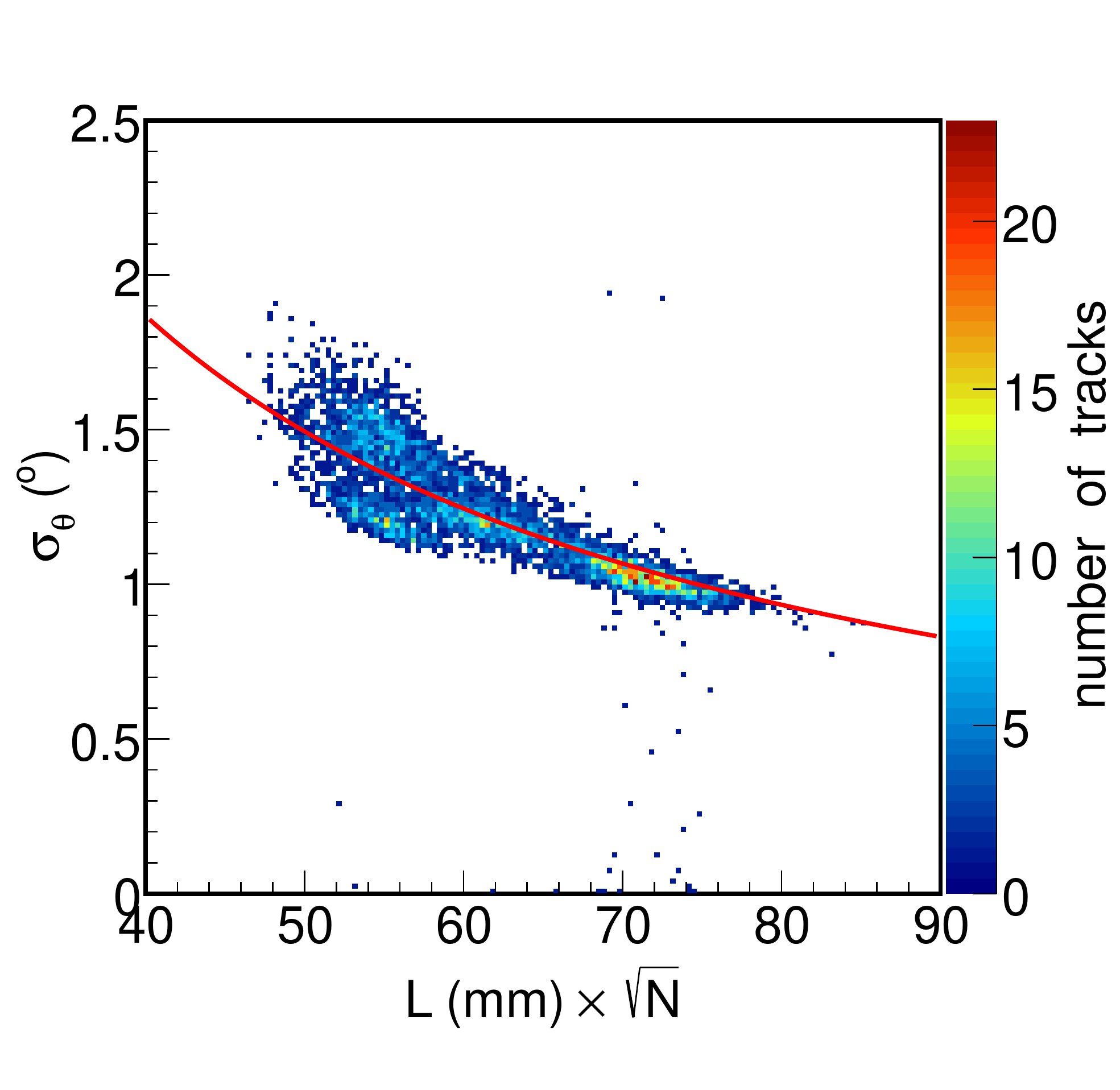}
\setlength{\abovecaptionskip}{0pt}
\caption{Track fitter uncertainties in $\phi$ (top) and $\theta$ (bottom) versus track length ($L$) and number of pixels hit ($N$) for alpha-particle events. The smooth curves are fits of Equation \ref{eq:res2} to the distributions.\label{fig:res_vs_npoints}}
\setlength{\belowcaptionskip}{0pt}
\end{figure}

Method III extracts the angular resolution for alpha tracks analytically. This allows the generalization of our results to a wider set of track types. Assuming the charge is deposited uniformly along a straight line (according to a SRIM simulation \cite{SRIM}, this is a good approximation for the short mid-section of the alpha-particle trajectories measured), it can be shown by a simple estimate that the track angle resolution depends on the point resolution as 
\begin{equation}
\begin{aligned}
 \sigma_{\phi,\theta}=\frac{\sqrt{12}~\sigma_{y,z}}{L\sqrt{N}} (\rm radians).
\label{eq:res2}
\end{aligned}
\end{equation}
Here, $\sigma_{\phi,\theta}$ is the point resolution in the coordinate direction that determines the angle of interest, $L$ is the track length, and $N$ is the number of space-points measured. The alpha tracks considered here traverse the whole chip width in $y$, so that $L \approx 7.2$~mm, and they are roughly parallel to the $x$-axis. Hence, when calculating $\sigma_\phi$, we use $\sigma_y=275~\mu {\rm m}$, and when calculating $\sigma_\theta$, we use $\sigma_z=313~\mu {\rm m}$. In order to verify the predicted dependence of the resolution on $N$ and $L$, we remove our selection on the impact parameter. This yields a set of tracks with a larger variation in the number of spacepoints, because the tracks traverse regions of the chip with different efficiencies. Figure~\ref{fig:res_vs_npoints} shows the resolution reported by the track fitter for these events. The red line in the figure shows the result of fitting Equation \ref{eq:res2} to the data points, where we allow $\sigma_{y,z}$ to float in the fit. The analytical expression appears to account correctly for the dependence of the resolution on track length and the number of space points measured. The point resolutions obtained from the fit are $\sigma_y=(332\pm1)~\mu{\rm m}$ and $\sigma_z=(376\pm 1)~\mu{\rm m}$, both 20\% higher than the actual point resolutions. While not a precise agreement, this means that the track angle resolution for a range of track lengths and number of measurement points can be predicted at the 20\%-level from the point resolution, which in turn can be predicted reliably from analytical estimates based on pixel and GEM feature sizes, as demonstrated in Section \ref{sec:pointres}. The present work thus provides a validated foundation for optimizing future, larger detectors, and for estimating the performance requirements on the components needed.

\section{3-D Tracking of Helium Nuclei: Energy Resolution}

\begin{figure}
\centering
\includegraphics[width=8cm]{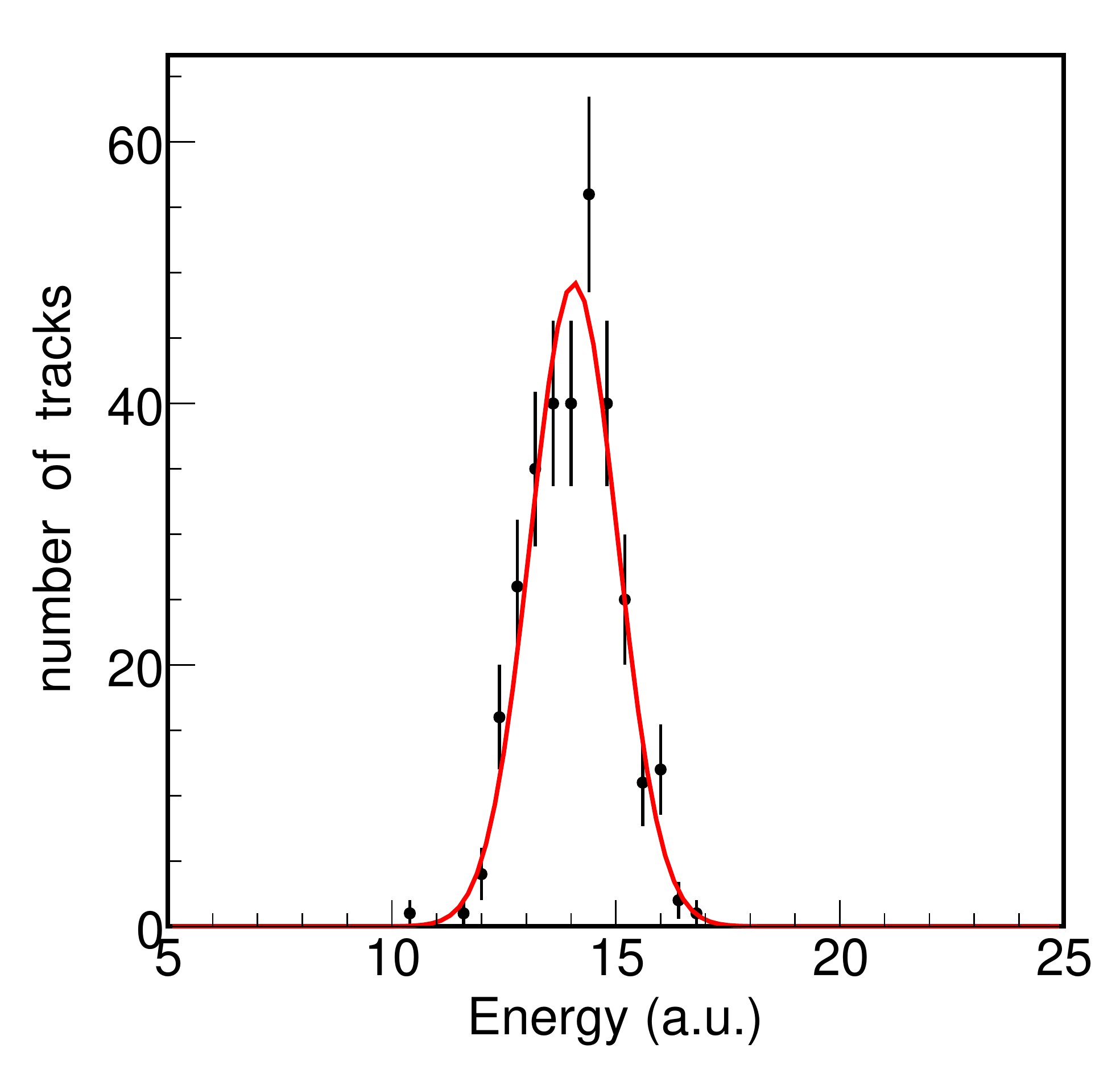}
\setlength{\abovecaptionskip}{0pt}
\caption{Reconstructed (uncalibrated) energy of alpha-particle events, recorded with a Po-210 source placed 3.88 cm from the detector, which was operating with HeCO$_2$ gas. The black points show experimental data, and the curve is a Gaussian fit.\label{fig:energy_resolution}}
\setlength{\belowcaptionskip}{0pt}
\end{figure}

\begin{figure*}
\centering
\includegraphics[width=8.5cm]{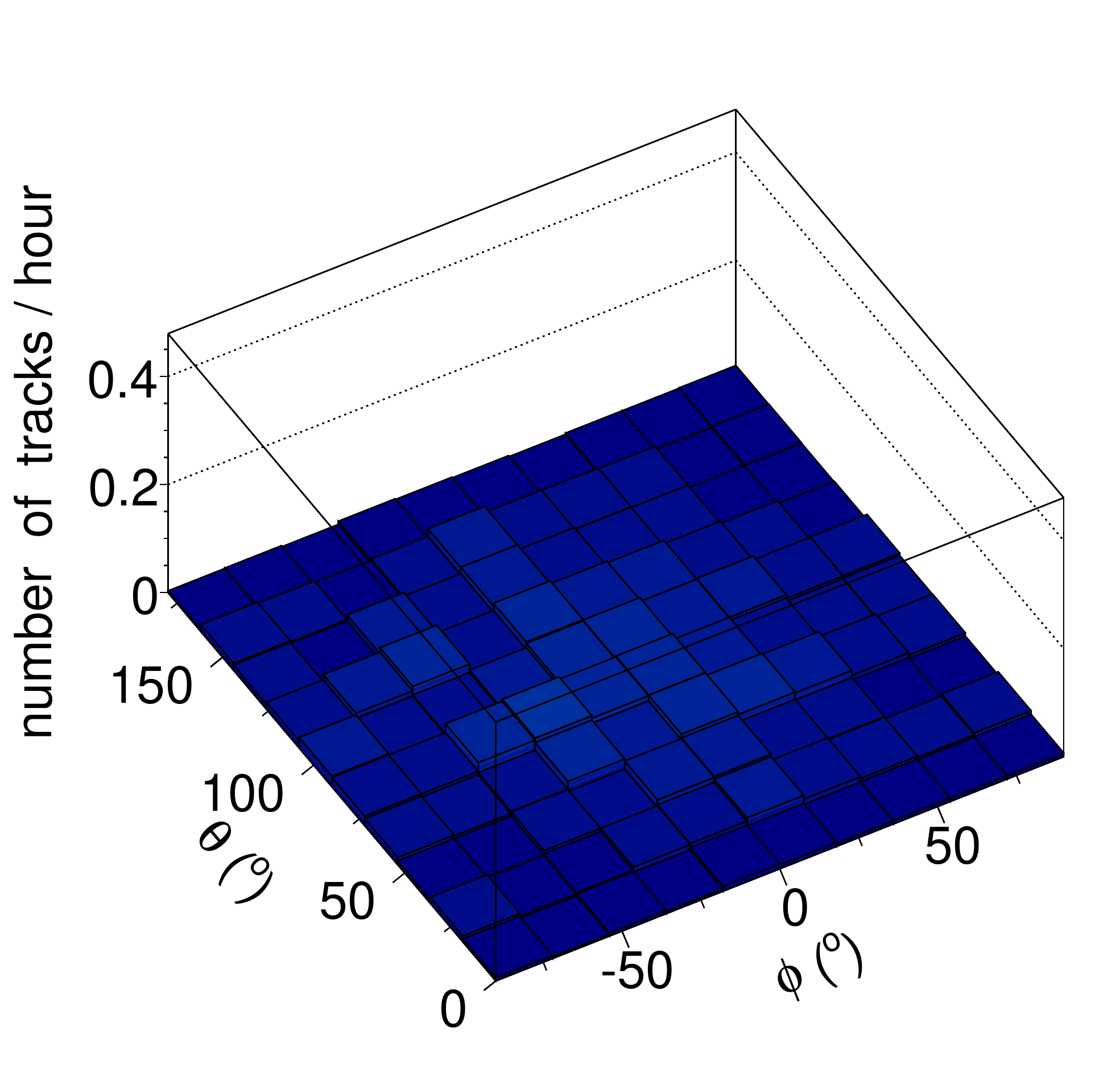}\includegraphics[width=8.5cm]{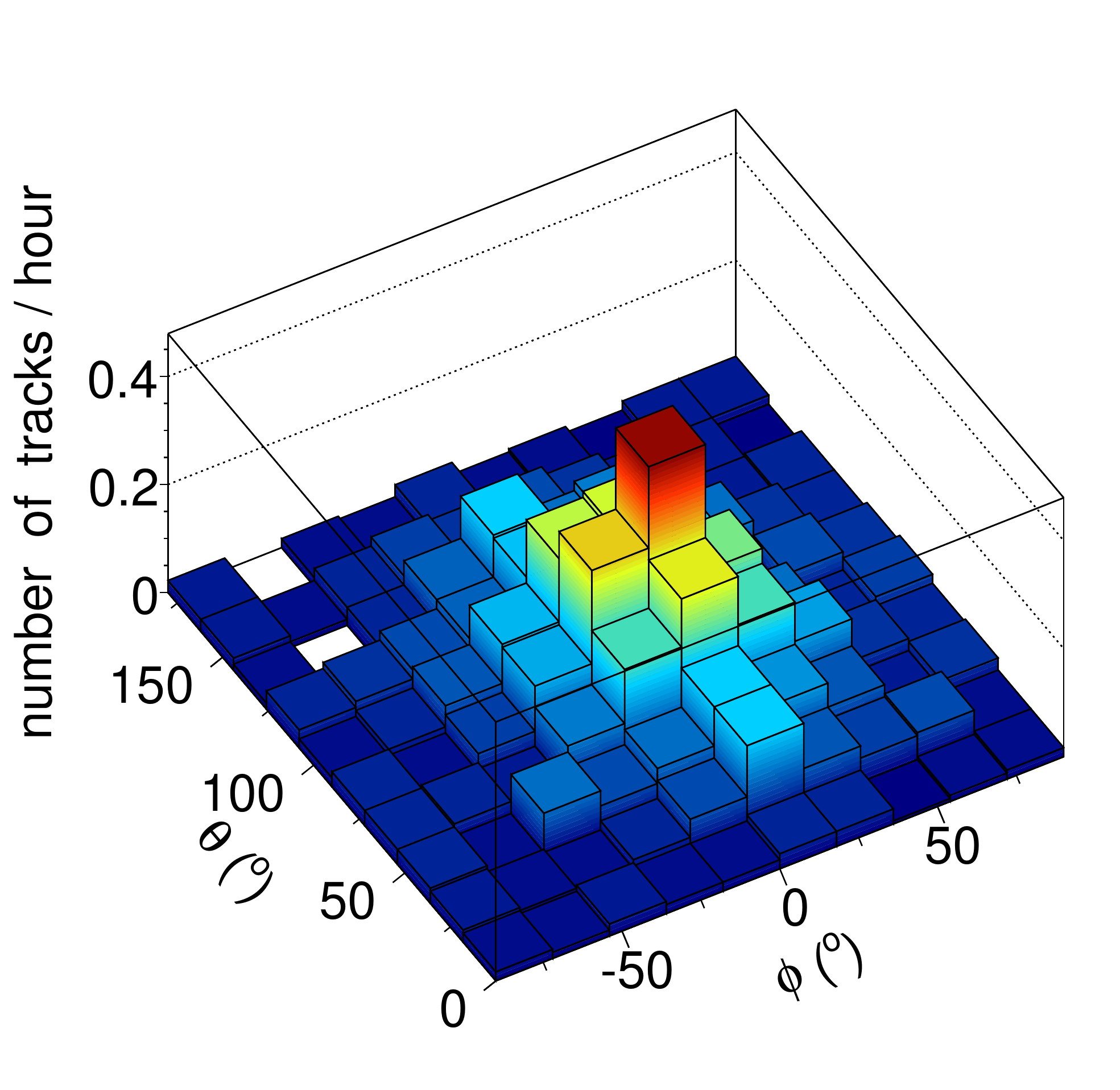}
\setlength{\abovecaptionskip}{0pt}
\caption{Time-normalized recoil angle distribution without (left) and with (right) a Cf-252 fast neutron source near the detector. The source is detected with high significance, and the recoils point back to the source.\label{fig:recoil_angles}}
\setlength{\belowcaptionskip}{0pt}
\end{figure*}

We estimate the fractional energy resolution of the detector for MeV-scale energies using the alpha particle sample discussed in the previous section. The energy of each reconstructed track is obtained by converting the time over threshold (ToT) measured by each pixel into an energy value, and adding up the energy values for all pixels in the event. The conversion from ToT to energy requires a calibration of the ToT measurement in each pixel, which is performed with a charge injection circuit internal to the pixel chip. The calibration requires knowledge of the gain, which was measured as described in section \ref{sec:gain}, and the work function $W$ for the gas. After performing this calibration procedure, the energy scale of the detector should be uniform across pixels. Contrary to expectation, however, we find that the energy scale varies strongly across the chip. Studies at LBNL \cite{kadyk} revealed that this is caused by imperfect contact between the chip and the conductive layer that was deposited onto it, so that at certain positions some of the GEM avalanche charge does not reach the pixel chip. We have already built an upgraded detector with the next generation (FE-I4) pixel chip, where we employed improved metal deposition, and this seems to have resolved the issue \cite{beasttpc}. Once we realized that the metal layer was the culprit, we gave up on an absolute energy scale calibration of the current detector, and we quote energy measurements here in uncalibrated, arbitrary units (a.u.). Figure~\ref{fig:energy_resolution} shows the energy measured with the pixel chip for the same alpha sample as was used to measure the angular resolution. A Gaussian fit determines the fractional energy resolution to be $\sigma_E/E=6.9\%$. The energy resolution is good, but quite a bit worse than the outstanding gain resolution, expected to be of order $2-3\%$ for MeV-scale signals and the gain used. The discrepancy appears to be due to the imperfect metallization. Also the energy resolution is improved in our next-generation detector \cite{beasttpc}.

\section{Directional Detection of Fast Neutrons \label{sec:neutrons}}

Having established the excellent performance of the \dcube prototype, we expose it to a $50~\mu $Ci Cf-252 fast neutron source. The average expected track length of reconstructed He-recoils from neutron-He elastic scattering is 1.6~mm at atmospheric pressure, which should be easy to detect, given the established detector performance. However, given the tiny detector volume and hence scattering probability, a very long exposure is required. Figure~\ref{fig:recoil_angles} shows the event rate versus 3-D recoil angle after selecting recoil-like events, for a 56-day source-free background run and a 7.5-day run with the neutron source 47 cm from the detector. The event rate is 27-sigma larger with the source present, the excess seen agrees with expectations from a detailed simulation \cite{Jaegle:2012sma}, and the observed recoil angle distribution points correctly back to the source. Clearly, already this tiny prototype can detect fast neutrons, and locate a point-like source in 3-D. More quantitative detail on the performance of a larger detector will be published soon.

\begin{figure}
\centering
\includegraphics[width=9cm]{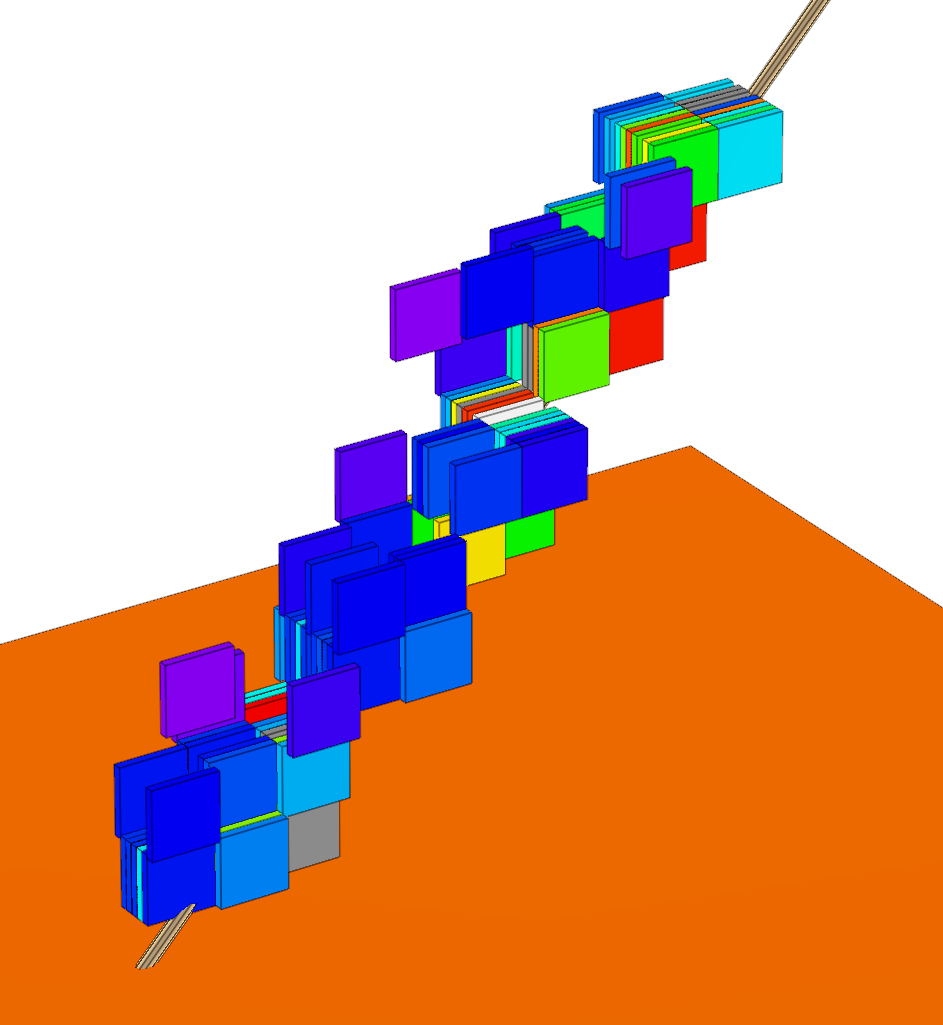}
\setlength{\abovecaptionskip}{0pt}
\caption{He-recoil candidate in HeCO$_2$ gas, measured with Cf-252 source near the detector.\label{fig:recoil}}
\setlength{\belowcaptionskip}{0pt}
\end{figure}

Note that the width of the recoil angle distribution is mainly due to scattering kinematics; the contribution from detector resolution (Section \ref{angles}) is negligible. Figure~\ref{fig:recoil} shows one of the neutron-recoil event candidates measured with the Cf-252 source present.

The seemingly minor mechanical modifications we made to the support structure for this particular study (Fig.~\ref{detector2}) were crucial for obtaining the clean directional signal. The larger detector volume increased the signal efficiency by a factor of five, while the reduced amount of plastic lowered the rate of the two main backgrounds, protons from neutron scattering in the plastic and alpha-particles not related to the neutron source (presumably from decaying radon-progeny stuck to the plastic) by an order of magnitude.

\section{Discussion of Results and Conclusion}

We demonstrated the 3-D reconstruction of mm-length alpha-track segments and nuclear recoils. We investigated two main performance measures of interest in that context, angular resolution and energy resolution, and how these are determined by choice of detector components, geometry, and gas mixture.

The point (single-hit) resolution of the detector can be reliably predicted from the GEM hole spacing, pixel segmentation, and diffusion in the different detector regions, and was found to be of order $200~\mu m \oplus~C \sqrt z$, where the first term is the resolution of the readout plane, $C$ is the diffusion per $\sqrt z$, $z$ is the drift length, and $\oplus$ denotes addition in quadrature. Interestingly, in the most precise coordinate, $y$, the resolution of the readout plane is no longer limited by detector segmentation, but rather diffusion in the transfer and collection gaps. We showed how the point resolution in turn determines the track angle resolution, and how this resolution scales with track parameters. For the mm-length alpha-track segments studied, the angular resolution was of order 1 degree. By combining equations \ref{eq:res} and \ref{eq:res2}, we find that the angular resolution in larger detectors will be approximately $\sqrt{12}(200~\mu m \oplus C \sqrt z)/(L \sqrt N)$, where $L$ is the track length, $N$ is the number of space points, and $C$ is the diffusion per $\sqrt z$. The implication is that tracks of length 5-10~mm can be reconstructed with angular resolution of order a few degrees in detectors with electron drift ($C\approx  200~\mu m/\sqrt{\rm cm}$) and short drift length ($z\lesssim 10~cm)$. For detectors with longer drift length or improved track angle resolution, negative ion drift \cite{Martoff:2000wi, Miyamoto:2004dc} would be advantageous to reduce the otherwise dominant contribution from diffusion in the drift gap.

The energy resolution measured for MeV-scale signals was $\sigma_E/E=7\%$, limited by a position dependence in the charge collection efficiency due to imperfect contact between the pixel chip and a metal layer that was deposited onto it. This has been resolved in the next generation detector \cite{beasttpc}. The measured asymptotic (high-gain) gain resolution, which we expect to limit the energy resolution in future detectors, varies from $\sigma_E/E=15\%$ at 3~keV, limited by statistical fluctuations in the primary ionization and in the gas avalanche process, to $\sigma_E/E=2\%$ at the MeV-scale, limited by detector and measurement stability. Improved gain resolution and thus energy resolution at low energies may be achieved by counting electrons individually, which again may require negative ion drift \cite{Sorensen:2012qc}.

The TPC charge readout technology under study looks promising for future directional WIMP searches and neutron detectors. The angular and energy resolutions measured are excellent, and the 3-D tracking capability reduces the number of signal events needed to claim a dark matter observation. The technology studied has other advantages, among them negligible noise rates and exceptionally high sensitivity. More precise studies are needed, but for now, the lack of noise hits in our event displays and the successful 3-D reconstruction of cosmic ray induced tracks with energies of order 1~keV demonstrate the potential for achieving directional detection with low track energy threshold. This will be important in the context of searching for the keV-scale nuclear recoil energies expected from the interaction of low-mass ($\approx$10 GeV) WIMPs. In the work presented we focused on MeV-scale nuclei and atmospheric gas pressure. We did not operate at low gas pressure ($\approx$10-50~Torr), which is required to extend the length of keV-scale recoils to measurable size \cite{Jaegle:2012sma}, nor did we use target gases optimized for WIMP sensitivity. We have already performed other work in this direction \cite{Vahsen:2014mca}. Further measurements with additional gas mixtures, e.g. CF$_4$, CF$_4$:CS$_2$ \cite{Daw:2010ud} and CF$_4$:CS$_2$:O$_2$ \cite{Snowden-Ifft:2014taa} at low gas pressures are required and planned. For low-mass WIMP searches, another outstanding question is the significance of straggling for keV-scale recoils, which we plan to measure. For now we can conclude that if such recoils deposit charge in the shape of tracks, then a detector based on GEMs and pixel ASICs should be able to reconstruct them in 3-D. As for neutron detection, we demonstrated that the technology under study can locate a fast neutron source in 3-D. We expect that the next generation detector, where the energy measurement has been improved \cite{beasttpc}, should be able to simultaneously locate and measure the neutron spectrum of a source.
 
\section{Acknowledgements}

We thank Marc Rosen for his assistance in designing the support structure and test vessel. We thank Blake Pollard and Kamaluoawaiku Beamer for performing electric field simulations. We acknowledge support from the U.S. Department of Homeland Security under Award Number 2011-DN-077-ARI050-03 and the U.S. Department of Energy under Award Number DE-SC0007852.

\section{References}
\bibliographystyle{elsarticle-num}

\end{document}